\PassOptionsToPackage{pdfpagelabels=false}{hyperref} 
\documentclass[useAMS,usenatbib]{mnras}
\usepackage[english]{babel}
\usepackage{hyperref}
\usepackage{graphicx}
\usepackage{float}
\usepackage{color}
\usepackage{multirow}
\usepackage{amssymb}
\usepackage{array}
\usepackage{placeins}
\usepackage[normalem]{ulem}
\usepackage{eso-pic}
\usepackage{siunitx}
\hypersetup{draft}

\AddToShipoutPictureBG*{%
  \AtPageUpperLeft{%
    \hspace{0.75\paperwidth}%
    \raisebox{-3.5\baselineskip}{%
      \makebox[0pt][l]{\textnormal{DES 2019-0048}}
}}}%

\AddToShipoutPictureBG*{%
  \AtPageUpperLeft{%
    \hspace{0.75\paperwidth}%
    \raisebox{-4.5\baselineskip}{%
      \makebox[0pt][l]{\textnormal{Fermilab PUB-19-114-AE}}
}}}%

\def\aap{Astronomy \& Astrophysics}
\def\aj{The Astronomical Journal}
\def\apj{The Astrophysical Journal}
\def\apjs{Astrophysical Journal Supplement Series}
\def\apjl{The Astrophysical Journal Letters}
\def\assp{Astrophysics and Space Science Proceedings}

\def\mnras{Monthly Notices of the Royal Astronomical Society}

\def\pasp{Publications of the Astronomical Society of the Pacific}
\def\pasj{Publications of the Astronomical Society of Japan}
\def\procspie{Proceedings of Society of Photo-Optical Instrumentation Engineers}

\def\nat{Nature}
\usepackage[switch]{lineno}
\newcommand{\diff}{\mbox{${\rm d}$}}

\newcommand{\Msun}{\mbox{M$_{\odot}$}}

\newcommand{\beq}{\begin{equation}}
\newcommand{\eeq}{\end{equation}}
\newcommand{\beqa}{\begin{eqnarray}}
\newcommand{\eeqa}{\end{eqnarray}}

\newcommand{\trilegal}{\textsc{Trilegal}}
\newcommand{\mwf}{\textsc{mwfitting}}

\title[Fitting MW using DES-Y3 data]{Modelling the Milky Way.
I - Method and first results \\
fitting the thick disk and halo with DES-Y3 data}

\author[Pieres, A., et al.]{A.~Pieres\thanks{E-mail: adriano.pieres@linea.gov.br}$^{1,2}$, L.~Girardi$^{1,3}$, E.~Balbinot$^{4}$, B.~Santiago$^{1,5}$, L.~N.~da Costa$^{1,2}$,
\newauthor
A.~Carnero~Rosell$^{1,6}$, A. B.~Pace$^{7}$, K.~Bechtol$^{8}$, M.~A.~T.~Groenewegen$^{9}$,
\newauthor
A.~Drlica-Wagner$^{10,11}$, T.~S.~Li$^{10,11}$, M.~A.~G.~Maia$^{1,2}$, R.~L.~C.~Ogando$^{1,2}$,
\newauthor
M.~dal Ponte$^{1,5}$, H.~T.~Diehl$^{10}$, A.~Amara$^{12}$, S.~Avila$^{13}$, E.~Bertin$^{14,15}$,
\newauthor
D.~Brooks$^{16}$, D.~L.~Burke$^{17,18}$, M.~Carrasco~Kind$^{19,20}$, J.~Carretero$^{21}$,
\newauthor
J.~De~Vicente$^{6}$, S.~Desai$^{22}$, T.~F.~Eifler$^{23,24}$, B.~Flaugher$^{10}$, P.~Fosalba$^{25,26}$,
\newauthor
J.~Frieman$^{10,11}$, J.~Garc\'ia-Bellido$^{13}$, E.~Gaztanaga$^{25,26}$, D.~W.~Gerdes$^{27,28}$,
\newauthor
D.~Gruen$^{17,18,29}$, R.~A.~Gruendl$^{19,20}$, J.~Gschwend$^{1,2}$, G.~Gutierrez$^{10}$,
\newauthor
D.~L.~Hollowood$^{30}$, K.~Honscheid$^{31,32}$, D.~J.~James$^{33}$, K.~Kuehn$^{34}$,
\newauthor
N.~Kuropatkin$^{10}$, J.~L.~Marshall$^{7}$, R.~Miquel$^{21,35}$, A.~A.~Plazas$^{36}$, E.~Sanchez$^{6}$,
\newauthor
S.~Serrano$^{25,26}$, I.~Sevilla-Noarbe$^{6}$, E.~Sheldon$^{37}$, M.~Smith$^{38}$, M.~Soares-Santos$^{39}$, 
\newauthor
F.~Sobreira$^{1,40}$, E.~Suchyta$^{41}$, M.~E.~C.~Swanson$^{20}$, G.~Tarle$^{28}$, D.~Thomas$^{42}$, 
\newauthor
V.~Vikram$^{43}$, A.~R.~Walker$^{44}$
\\
Affiliations are listed after the references
}

\begin{document}

\date{Accepted in XX YY 2019. Received 08 April 2019}

\pagerange{\pageref{firstpage}--\pageref{lastpage}} \pubyear{2019}

\maketitle

\label{firstpage}

\begin{abstract}
We present a technique to fit the stellar components of the Galaxy by
comparing Hess Diagrams (HDs) generated from \trilegal\  models to real data. We apply this technique, which we call \mwf , to photometric data from the first three years of the Dark Energy Survey (DES).
After removing regions containing known resolved stellar systems such as globular clusters, dwarf galaxies, nearby galaxies, the Large Magellanic Cloud and the Sagittarius Stream, our main sample spans a total area of $\sim$2,300 deg$^2$. 
We further explore a smaller subset ($\sim$ 1,300 deg$^2$) that excludes all regions with known stellar streams and stellar overdensities.
Validation tests on synthetic data possessing similar properties to the DES data show that the method is able to recover input parameters with a precision better than 3\%.
We fit the DES data with an exponential thick disk model and an oblate double power-law halo model.  We find that the best-fit thick disk model has radial and vertical scale heights of $2.67\pm0.09$ kpc and $925\pm40$ pc, respectively. The stellar halo is fit with a broken power-law density profile with an oblateness of $0.75\pm0.01$, an inner index of $1.82\pm0.08$, an outer index of 4.14$\pm$0.05, and a break at $18.52\pm0.27$ kpc from the Galactic center.
Several previously discovered stellar over-densities are recovered in the residual stellar density map, showing the reliability of \mwf\  in determining the Galactic components. 
Simulations made with the best-fitting parameters are a promising way to predict MW star counts for surveys such as the LSST and Euclid.
\end{abstract}


\section{Introduction}
\label{sec:intro}

Over the last 40 years, we have learned  the utility of describing a complex system like the Milky Way (MW) through simple building blocks \citep[e.g.,][]{1981ApJS...47..357B}, composed of nearly homogeneous stellar populations, smoothly distributed in space in a few components like the thin and thick disks, bulge and halo.
The derivation of simple parameters for these components -- such as scale lengths and heights, limiting radii, central densities, etc. -- allows us to put our Galaxy in perspective by comparing it to other spiral galaxies \citep{2011ApJ...739...20C} and to  galaxies produced in cosmological simulations \citep[see, e.g.][]{2014MNRAS.445..581H,2016ARA&A..54..529B}.  
Examining the residuals of the best-fit models enables the identification of stellar substructure such as dwarf galaxies and stellar streams \cite[e.g.,][]{2018AAS...23121205S}. Fitted models can also be used to estimate the distribution of stars in future surveys.

Our understanding of the MW has steadily advanced over the past several decades. For example, the thick disk \citep{1983MNRAS.202.1025G} has long been proposed to explain the MW stellar population within 1-5~kpc on either side of the Galactic plane. 
Thick disk stars differ from those closer to the Galactic plane in kinematics, age and metalicity, being older, more metal-poor, less rotationally supported, and having typically higher [$\alpha$/Fe] at a fixed metalicity \citep[for instance, see][]{2006MNRAS.367.1329R, 2008MNRAS.384..173F}. More recently, the spatial structure of different stellar populations has been studied by \cite{2014A&A...564A.115A} and \cite{2016ApJ...823...30B}, among others, using survey data from APOGEE \citep{2016AN....337..863M}. In brief, high [$\alpha$/Fe] stars tend to follow a double exponential density profile parallel and perpendicular to the Galactic plane, with scales of $h_R \simeq 2.2$~kpc and $h_z \simeq 1.0$~kpc, respectively \citep{2016ApJ...823...30B}. The lower [$\alpha$/Fe] stars display a more complex distribution, including a metalicity gradient and disk flaring~\citep{2014A&A...564A.115A}. Even so, the traditional description of the thin and thick disk components with double exponential profiles (or a sech$^2 z$ perpendicular to the disk plane) is  adequate~\citep{2005A&A...433..173C,2008ApJ...673..864J, 2010ApJ...714..663D}. 

At the outer limit of the MW, the stellar Galactic halo is roughly spherical in shape. Early studies indicated that the radial density of this component is better described by a power-law profile with index $n \sim -2.75$ than an exponential profile \citep{2008ApJ...673..864J,2010ApJ...714..663D}.
However, more recent work has found that the stellar density drops off faster at typical distances $\simeq$ 20 kpc, suggesting that the density of the stellar halo follows a broken power-law profile
\citep{2009MNRAS.398.1757W,2011MNRAS.416.2903D,2011ApJ...731....4S,2018ApJ...852..118D} or another model that decreases more rapidly at large radii \citep{1965TrAlm...5...87E,2006AJ....132.2685M,2011MNRAS.416.2903D,2018ApJ...859...31H}. These observations are not unexpected, since a power-law index of $n < -3$ is necessary at large radii in order for the integrated mass of the stellar halo to converge.

In addition to the aforementioned developments in describing the stellar content of the Galaxy, an impressive amount of work has been dedicated to determine the star formation rate \citep[SFR, ][]{1991AJ....101.1865R, 1998A&A...338..161F}, initial mass function \citep[IMF, ][]{2001MNRAS.322..231K, 2003PASP..115..763C, 2003ApJ...598.1076K, 2005MNRAS.362..945W}, and Age-Metalicity Relation \citep[AMR,][]{2000A&A...358..850R, 2003A&A...399..931Z, 2008MNRAS.384..173F, 2011A&A...530A.138C} for the stars in the MW, along with the modelling of stellar evolution \citep{1994A&AS..106..275B, 2000A&AS..141..371G, 2002A&A...391..195G, 2006ApJS..162..375V, 2007A&A...469..239M, 2010ApJ...724.1030G, 2011ApJS..192....3P, 2013ApJ...776...87S} and the stellar contents of the Galaxy itself (\citealt{2011ApJ...730....3S}; \citealt{2014A&A...564A.102C}\footnote{See \url{https://model.obs-besancon.fr/modele_ref.php} for a complete list of publications of the Besan\c{c}on group.}; \citealt{2018ApJ...860..120P}). Thanks to all these developments, we are now able to build a detailed structural model for the Galaxy.

To take advantage of this knowledge and the increasing number of deep wide-field astronomical surveys, we have  developed \mwf. This work aims to present the method and to show its first application to data in the Dark Energy Survey (DES; \citealt{DES2005}). 
 
In this work we aim to:
\begin{itemize}
\item Present an efficient method to describe the structure of the Galaxy  by comparing star counts to predictions of stellar population synthesis models. The comparison between data and models is made through  binned colour-magnitude diagrams (i.e., Hess Diagram, HD) in  specific regions in the sky. Many different models are used to predict star counts, such as the spatial distribution of stars in the MW components, the stellar IMF, SFR, and AMR. Also crucial in determining  star counts are the input stellar evolutionary models that prescribe magnitudes and colours as a function of fundamental stellar parameters, such as mass, age, and metalicity.

\item Validate the code using mock data. 
These tests are done to test the accuracy of \mwf\, to evaluate systematic uncertainties, and to measure the effect initial values has on recovering the input parameters.

\item Apply \mwf\ to model the Galactic thick disk and halo in DES three year (Y3) data. 

\item Show and discuss the results of the method and the implications on the Galactic model adopted.
\end{itemize}

This paper is structured as follows: in Section~\ref{sec:parameters} we discuss the \mwf. In Section~\ref{sec:data} we briefly describe the DES Y3 data. In Section~\ref{sec:results} we present the results of \mwf. In Section~\ref{sec:sim} we describe a  simulation based on the best fitting parameters and discussion of the results. Finally, we conclude in  Section~\ref{sec:conc}.

	\section{\textsc{MWFitting} package}
	\label{sec:parameters}

In this paper, we adopt \trilegal\footnote{\url{http://stev.oapd.inaf.it/cgi-bin/trilegal}} models to describe the stellar content of the Galaxy. \trilegal\ is a stellar population synthesis code, based on the \citet{2002A&A...391..195G} database of stellar isochrones, and augmented with models for brown and white dwarfs. For more details about the stellar models, we refer to \citet{2005A&A...436..895G}. Note that even though several upgrades in the database of evolutionary tracks and stellar atmospheres have become available recently \citep[see, e.g.][]{2017ApJ...835...77M}, they severely reduce computational speed, and only include short-lived evolutionary phases and cool stars, which are not the subject of the present work.

The following subsections present the sequence of steps that leads to a final product of the \mwf\ . Section~\ref{tripar} describes  \trilegal\ input parameters to model a sky region with a specific Galactic model. The previous attempts to calibrate the Galactic model using \trilegal\ are briefly discussed in Section~\ref{pretri}; the adopted Galactic model is presented in Section~\ref{sec:galmod}; in Section~\ref{mwfcode} we discuss the implementation of the \mwf; in Section~\ref{validation} we validate the \mwf\ pipeline using synthetic data with known input parameters. 
 
\subsection{\trilegal\ parameters}
\label{tripar}

The \trilegal\ population synthesis simulation requires input parameters such as: covered area, photometric system, magnitudes and colour ranges, 3D position of the Sun, dust distribution, IMF for single stars,  binary fraction, and mass ratios of unresolved binaries. It also requires  structural models, SFR, and AMR for each Galactic component (see Table \ref{tabparameters}). 

Regarding the color and magnitude ranges, \trilegal\ models are very successful in describing the stellar evolutionary phases as the main sequence (MS), including the turn-off (MSTO), and stars in the sub and red giant branches (respectively, SGB and RGB), for stars in a wide range of masses.

Stellar evolutionary models present a poor colour-fit for low-mass stars with [Fe/H] $\geq$ -2, such as M-type stars, which is the most abundant spectral type in thin disk. See for instance \citep{2007AJ....133.1658S}, for a discussion about the comparisons of simple stellar populations of globular clusters to theoretical models. 

Based on that, we choose to exclude the red thin-disk stars (see figure 2 and discussion in \citealt{2010ApJ...714..663D}) and keep the parameters of this component fixed. The magnitude depth of DES also favours stars farther away than those in the thin disk, which supports our choice.

\subsection{Previous attempts to calibrate \trilegal}
\label{pretri}

Early descriptions of the MW components and their calibrations using \trilegal\ are found in \citet{2002A&A...392..741G} and \citet{2005A&A...436..895G}. Those first attempts were based on a simple trial-and-error approach, where each model parameter was set to literature values, changed by hand until a ``good description'' for the star counts was met for a given survey. Surveys used in these analyses compromise both deep \citep[e.g., DMS and EIS-deep,][]{1998ApJS..119..189O, 2001A&A...379..740A},  shallow  \citep[e.g., ,][]{2006AJ....131.1163S} photometric data, and  local  \citep[e.g., \textit{Hipparcos} catalogue,][]{1997A&A...323L..49P}.

\citet{2009A&A...498...95V} explored a different approach to calibrate the bulge's parameters using \trilegal. They defined a likelihood function to quantitatively evaluate the goodness-of-fit between data and model \citep[see also ][]{2004PhLB..592....1P, 2002MNRAS.332...91D} as:
\begin{equation}
\label{like}
-2 \ln \lambda(\theta) = 2 \sum_{i=1}^{N}
\left( \nu_i(\theta)-n_i+n_i\ln\frac{n_i}{\nu_i(\theta)} \right)
\end{equation}
where $n_i$ is the number of observed objects in a given magnitude/colour bin \emph{i}, and $\nu_i(\theta)$ is the number of objects predicted by the set of parameters $\theta$ that describes the model. The summation is performed over all lines-of-sight, and magnitude/colour bins included in the comparison. The authors used the Broyden-Fletcher-Goldfarb-Shanno (BFGS) algorithm \citep{1987Prmo.book.....F} to maximize their likelihood and derived uncertainties from the likelihood profile, as detailed in that work.

In this context, the fitting of disk and halo parameters using the latter method requires an extra set of variables. This presents several issues:
\begin{itemize}
\item Fitting the disk (thin and thick) and halo implies the simultaneous fitting of $\sim30$ structural parameters, with many samples across the sky. The resulting analysis is very time-consuming.

\item Local maxima in likelihood space may be very common, and due to the high dimensionality of the problem, finding the absolute maximum may be challenging.

This is not the case when fitting the bulge, as there are fewer parameters, and there are a large set of lines-of-sight which leaves little chance for solutions to be trapped in local maxima \citep{2009A&A...498...95V}.
In the present case, it is advisable to implement tests for local maxima in log-likelihood space, and  check whether different starting conditions lead to the same solution. These tests imply even longer computing times. 
\end{itemize}

In the following sections, we describe the  implementation of \mwf\ aimed at tackling the challenges discussed above \citep[see also][]{2012ASSP...26..165G}.

\begin{table*}
\begin{center}
\renewcommand{\arraystretch}{1.06}
\caption{The MW model adopted in this work includes the bulge (as a triaxial truncated spheroid component), the thin disk (as an exponential model in the radial direction and a squared hyperbolic secant model in the vertical direction), the thick disk (as an exponential model in both $R$ and $z$ directions), and the halo (modeled with a double power-law profile). The columns list: the formula describing each MW component (first), free parameters (second), a description of each component (third), units (forth), initial value (fifth), and the best-fit value with errors (last column) for both samples ($^{\dagger}$ for \emph{raw} sample and $\ddagger$ for \emph{refined} sample).}
\label{tabparameters}

\setlength{\tabcolsep}{2.6pt}
\begin{tabular}{lllllr}
\hline
Formula & Symbol & Meaning & Unit & Initial value & Fixed/Best-fit value$^{\star}$ \\
\hline
\multicolumn{6}{c}{\textbf{Bulge}$^{1}$} \\
\vspace{-0.1cm}
\multirow{7}{*}{
\begin{minipage}{5.6cm}
\begin{eqnarray}
\rho^{\rm bulge} = \rho^{\rm bulge}_{GC} \frac{\exp(-a^2/a_m^2)}{(1+a/a_0)^{1.8}} \nonumber \\
\mbox{with } \rho^{\rm bulge}(0,0,0) = \rho^{\rm bulge}_{\rm GC} \nonumber \\
\mbox{with } a=\left( x'^2 + y'^2/\eta^2 + z^2/\zeta^2 \right)^{1/2}
\nonumber \\
\mbox{and } {x',y'} \mbox{ rotated by $\phi_0$. w.r.t. } x,y \nonumber
\end{eqnarray}
\end{minipage} 
}
 & & & & \\
 & $\rho^{\rm bulge}_{\rm GC}$ & space density at GC & \Msun\,pc$^{-3}$ & 400 & fixed \\ 
 & $a_{\rm m}$ & scale length & pc & 2500 & fixed \\
 & $a_{\rm 0}$ & truncation scale length & pc & 95 & fixed \\
 & $\eta$, $\zeta$ & 1:$\eta$:$\zeta$ scale ratios & - & 0.68, 0.31 & fixed \\
 & $\phi_0$ & angle w.r.t. Sun-GC line & deg ($^\circ$) & 15 & fixed \\
\\ \hline
\multicolumn{6}{c}{\textbf{Thin disk}} \\
\vspace{0.0cm} \\
\multirow{7}{*}{
\begin{minipage}{5.4cm}
\begin{eqnarray}\rho^{\rm thin}= A^{\rm thin}\mbox{sech}^2(h/h_z^{\rm thin}) \times \nonumber \\
\exp(R/h_R^{\rm thin})\nonumber \\ 
\mbox{with } h_z^{\rm thin}=h_{z,0}^{\rm thin}  + \left(1+t/t_{\rm incr}^{\rm thin}\right)^\alpha\nonumber\\ 
\mbox{and } \left.\int_{h=-\infty}^{+\infty}\rho^{\rm thin}\diff z\right|_\odot=\Sigma_\odot^{\rm thin}\nonumber
\end{eqnarray}
\end{minipage} 
}
 & $\Sigma_\odot^{\rm thin}$ & local mass surface density & \Msun\,pc$^{-2}$ & 55.41$^{2}$ & fixed \\ 
 & $h_R^{\rm thin}$ & thin disk scale length & pc & 2913$^{2}$  & fixed \\
 & $R_{\rm max}^{\rm thin}$ & maximum radius & kpc & 15 & fixed \\
 & $h_{z,0}^{\rm thin}$ & scale height for & pc & 94.7$^{2}$ & fixed \\
  &  & youngest stars &  &  \\
 & $t_{\rm incr}^{\rm thin}$ & timescale for increase in $h_z$ & Gyr & 5.55$^{2}$ & fixed \\
 & $\alpha$ & exponent for increase in $h_z$ & - & 1.67$^{3}$ & fixed \\
\hline
\multicolumn{6}{c}{\textbf{Thick disk}} \\
\vspace{0.0cm} \\
\multirow{7}{*}{
\begin{minipage}{5.4cm}
\begin{eqnarray}\rho^{\rm thick}= A^{\rm thick}\exp(h/h_z^{thick}) \times \nonumber \\
\exp(R/h_R^{\rm thick}) \nonumber \\ 
\mbox{with } \left.\int_{h=-\infty}^{+\infty}\rho^{\rm thick}\diff z\right|_\odot=\Sigma_\odot^{\rm thick}\nonumber
\end{eqnarray}
\end{minipage} 
}
 & $h_{z}^{\rm thick}$ & scale height & pc & 925 & $925\pm40^{\dagger}$ \\
 &  &  &  &  & $910\pm46^{\ddagger}$ \\
 & $h_R^{\rm thick}$ & thick disk scale length & pc & 2667 & $2667\pm95^{\dagger}$ \\
 &  &  &  & & $2631\pm121^{\ddagger}$ \\
 & $\Sigma_\odot^{\rm thick}$ & local mass surface density & $10^{-3}$\Msun\,pc$^{-2}$ & 3.89 & $3.89\pm0.65^{\dagger}$ \\ 
 & & & & & $3.97\pm0.74^{\ddagger}$ \\ 
 & $R_{\rm max}^{\rm thick}$ & maximum radius (fixed) & kpc & 15 & fixed \\
\hline
\multicolumn{6}{c}{\textbf{Halo}} \\
\vspace{0.0cm} \\
\multirow{7}{*}{
\begin{minipage}{5.4cm}
\begin{eqnarray} \rho^{\rm halo}= f \rho^{halo}_{\odot} \left(\frac{r_{\odot}}{r_{obl}}\right)^{n} \nonumber \\
\mbox{with } \rho^{halo}(R_{\odot},0,z_{\odot}) = \rho^{halo}_{\odot}, \nonumber \\
r_{obl} = \sqrt{R^2+(z/q^2)} \nonumber \\
\mbox{ if } r_{obl} \leq r_{br}, n=n_1, f=1   \nonumber \\
\mbox{ else } (r_{obl}>r_{br}), n=n_2, f=(r_{\odot}/r_{br})^{n_1-n_2} \nonumber \\
\nonumber \\
\nonumber \\
\nonumber
\end{eqnarray}
\end{minipage} 
}
& $n_1$ & inner exponent & - & 1.82 & $1.82\pm0.08^{\dagger}$ \\
& & & & & $1.86\pm0.11^{\ddagger}$ \\
& $n_2$ & outer exponent & - & 4.14 & $4.14\pm0.05^{\dagger}$ \\
& & & & & $4.24\pm0.06^{\ddagger}$ \\
& $r_{BR}$ & break radius & kpc & 18.52 & $18.52\pm0.27^{\dagger}$ \\
& & & & & $18.59\pm0.49^{\ddagger}$ \\
& $q$ & axial ratio $z/x$ & - & 0.75 & $0.75\pm0.01^{\dagger}$ \\
& & (oblateness) &  & & $0.74\pm0.02^{\ddagger}$ \\
& $\rho^{halo}_{\odot}$ & local mass space density & $10^{-5} \Msun\,$pc$ ^{-3}$ & 3.31 & $3.31\pm0.20^{\dagger}$ \\ 
& & & & & $3.51\pm0.26^{\ddagger}$ \\ 
\hline
\multicolumn{6}{c}{\textbf{Dust layer}}\\
\multirow{4}{*}{
\begin{minipage}{5.4cm}
\begin{eqnarray}\rho^{\rm dust}= A^{\rm dust}\exp(h/h_z^{\rm dust}) \nonumber \\ 
\mbox{with } \int_{\ell=0}^{+\infty}\rho^{\rm dust}\diff \ell=A_V^{\infty}\nonumber
\end{eqnarray}
\end{minipage} 
}
\\
 & $A_V^{\infty}$ & total extinction at infinity & - & $^4$ & fixed \\
 & $h_z^{\rm dust}$ & dust scale height & pc & 110$^{5}$ & fixed 
\vspace{0.5cm} \\ \hline
\multicolumn{6}{c}{\textbf{Others}} \\
\\
 & $R_\odot$ & Sun's distance to the GC & kpc & 8.122$^{6}$ & fixed \\
 & $z_\odot$ & Sun's height above plane & pc & 20.8$^{7}$ & fixed \\
\hline \\
\multicolumn{4}{l}{$^{1}$ Parameters from
{\cite{2009A&A...498...95V}}} & & \\
\multicolumn{4}{l}{$^{2}$ Best-fit parameter from {\cite{2005A&A...436..895G}}} & & \\
\multicolumn{4}{l}{$^{3}$ Adopted in {\cite{2005A&A...436..895G}}} & & \\
\multicolumn{4}{l}{$^{4}$ {\cite{1998ApJ...500..525S}}}  & & \\
\multicolumn{4}{l}{$^{5}$ {\cite{1982A&A...109..213L}}} & & \\
\multicolumn{4}{l}{$^{6}$
{\cite{2018A&A...615L..15G}}} & & \\
\multicolumn{4}{l}{$^{7}$ \cite{2019MNRAS.482.1417B}} & & \\
\multicolumn{4}{l}{$^{\star}$ See Table~\ref{finaltable} for more details about those parameters.} & & \\

\end{tabular}
\end{center}
\end{table*}

\subsection{Galactic model adopted}
\label{sec:galmod}

Table \ref{tabparameters} summarizes the functional form utilized for each Galactic component, the parameters that describe the component, and whether the parameter is fixed or free in the fit.
We adopt an exponential model along the disk plane and a square hyperbolic secant perpendicular to it for the thin disk. The parameters of the thin disk and bulge modelled by \trilegal\ in this work are kept fixed at the values described in \citet{2005A&A...436..895G}, with some minor tweaks as in \citet{2012ASSP...26..165G}. The only parameters allowed to vary are related to the thick disk and to the halo of our Galaxy. An exponential model in both radial and vertical directions describes the distribution of stars in the thick disk. 
The stellar halo is described by a double power-law profile, with an inner exponent, $n_1$, describing the stellar density of the halo out to a certain distance, $r_{BR}$ (radius of the break) and an outer exponent, $n_2$, for farther distances. We require that the density of halo stars is continuous at $r_{BR}$ for both exponents.
Since DES covers the south Galactic cap ($b<-30^{\circ}$), it largely excludes the MW bulge. Therefore, in this analysis, we fix the parameters of the bulge component following the triaxial model presented in \cite{1997MNRAS.288..365B}.

The IMF assumed for Galactic stars is the Chabrier lognormal IMF \citep{2003PASP..115..763C} and the fraction of binaries adopted is 30\%, with the mass ratio of the secondary over the primary limited between 0.7 and 1.0 \citep{2002A&A...385..847B}. The SFR and AMR are specific to each MW component. Stars in the bulge and in the thick disk follow a SFR and AMR described by \cite{2003A&A...399..931Z} and \cite{2013A&A...559A..59B}, respectively. Thin disk and halo stars are modelled following previous comparisons from \cite{2002A&A...392..741G} and \cite{2005A&A...436..895G}.

\subsection{The \textsc{MWFitting} pipeline: fitting the galaxy with Hess diagrams}
\label{mwfcode}

The \mwf\ pipeline code fits a global, multi-component
model of the MW to the observed stellar density in bins of Galactic
longitude and latitude, magnitude, and color. The inclusion of spatial
and color-magnitude information allows us to break degeneracies
between the various MW model components.

\begin{table*}
\begin{center}
\def\arraystretch{2}
\setlength{\tabcolsep}{4pt}
\caption{Results of two tests (A and B) using \mwf\,. Even though the initial guesses start far from the input values, the final parameter values are within 3\% of the input values. The simulations in this table compare 100 fields and oversample the models in the same way as the comparison to real data.}.

\begin{tabular}{|l|l|l|c|c|c|c|c|c|c|c|}
\hline
\textit{Parameter} & \textit{Unit} & \textit{True} & \multicolumn{2}{c|}{\textit{Initial Guess}} & \multicolumn{2}{c|}{\textit{Best-fitting}} & \multicolumn{2}{c|}{{$\frac{\mid Best-True \mid}{True}$} (\%)} & \multicolumn{2}{c|}{{$\frac{Best-True}{\sigma}$}} \\
\cline{4-5} \cline{6-7} \cline{8-9} \cline{10-11}
 &  & \textit{Value} & A & B & A & B & $\ \ \ \ $A$\ \ \ \ $ & B & $\ \ $A$\ \ \ $ & B \\
\hline
ThickDisk $h_z$ & $\mathrm{pc}$ & 925 & 1037.6 & 903.6 & $923.1^{+2.3}_{-1.9}$ & $922.5^{+1.9}_{-1.9}$ & 0.2 & 0.2 & -0.8 & -1.2 \\

ThickDisk $R_e$ & $\mathrm{pc}$ & 2666 & 2849 & 2397 & $2657^{+6}_{-6}$ & $2675^{+6}_{-6}$ & 0.3 & 0.3 & -1.5 & +1.5 \\

ThickDisk $\rho$ (R=R$_{0}$) & $\times 10^{-3}$M$_\odot\mathrm{pc}^{-2}$ & 3.90 & 4.23 & 3.57 & $3.91^{+0.02}_{-0.03}$ & $3.94^{+0.02}_{-0.02}$ & 0.3 & 1.0 & +0.4 & +1.7 \\

Halo $n_1$ & - & 1.821 & 1.501 & 1.991 & $1.861^{+0.010}_{-0.011}$ & $1.867^{+0.009}_{-0.010}$ & 2.2 & 2.5 & +3.8 & +4.8 \\

Halo $n_2$ & - & 4.139 & 4.407 & 4.520 & $4.133^{+0.005}_{-0.005}$ & $4.124^{+0.005}_{-0.005}$ & 0.1 & 0.4 & -1.2 & -3.0 \\

Halo $r_{br}$ & kpc & 18.52 & 17.61 & 20.06 & $18.65^{+0.04}_{-0.04}$ & $18.63^{+0.04}_{-0.05}$ & 0.7 & 0.6 & +3.0 & +2.5 \\

Halo $q$ & - & 0.748 & 0.785 & 0.827 & $0.745^{+0.001}_{-0.001}$ & $0.748^{+0.001}_{-0.001}$ & 0.4 & 0.0 & -3.0 & 0.0 \\

Halo $\rho$ (R=R$_{0}$) & $\times 10^{-5}$ M$_\odot\mathrm{pc}^{-3}$ & 3.31 & 3.36 & 3.52 & $3.40^{+0.02}_{-0.02}$ & $3.39^{+0.02}_{-0.02}$ & 2.7 & 2.4 & +4.5 & +4.0 \\
\hline 
\end{tabular}
\label{sim}
\end{center}
\end{table*}

We begin by pixelizing the sky using the \textsc{HEALPix}\footnote{\url{https://healpix.jpl.nasa.gov/}} scheme to define individual lines-of-sight (which we call ``cells"). We select cells that reside within the survey, and remove cells that are contaminated by resolved stellar populations such as globular clusters and dwarf galaxies. 
For each cell, we calculate the coordinates of the centre, the average reddening and reddening dispersion, the limiting magnitude (as specified by the user), the colour range, and the bin size in the color-magnitude diagram (CMD) space.

Within each cell, we calculate model HDs for each component (i.e., bulge, halo, thin, and thick disk) over a range of distance moduli, typically separated by 0.1 mag. These so-called ``partial HDs'' for each component and distance are stored in separate Header Data Units (HDUs) of a multi-extension FITS\footnote{\url{https:////fits.gsfc.nasa.gov/fits_standard.html}} file. This data format allows the normalizations of different model components to be quickly adjusted. For example, the normalization of the stellar halo can be adjusted by a factor $f$, by multiplying all partial HDs associated with the halo by the same factor $f$.  The total model-predicted MW HD can be quickly calculated from a linear combination of the individual partial HDs, incorporating typical photometric errors of the survey in each band. This method allows us to rapidly construct stellar density predictions for a wide range of MW model parameters as listed in Table 1. Variation in each parameter corresponds to varying the weight of each partial HD, which are then combined to produce a total HD in each \textsc{HEALPix} cell.

The Poisson log-likelihood (Eq. \ref{like}) is calculated by first comparing the total model-predicted HDs to the data in each cell and then summing the log-likelihoods over all cells. To fit the MW model to an observed data set, we apply an Affine Invariant Markov Chain Monte Carlo (MCMC) Ensemble sampler \citep[i.e.,\textsc{emcee},][]{2013PASP..125..306F}. The free and fixed parameters of our model, along with their initial values, are listed in Table~\ref{tabparameters}. We assume flat priors ranging from 0.5-2.0 times the initial value of each free model parameter.
We also checked visually whether the walkers converged or not at the end of the burn-in phase, in order to inform realistic best-fitting parameters.

Since \trilegal\ computes a discrete distribution of points as a
realization of the expected population of stars in each cell, we are left with statistical noise due to sampling a finite number of points. To mitigate this noise, we increase the number of simulated stars by an \emph{over-factor} which is then taken into account while normalizing the final Hess diagram for each cell. A typical \emph{over-factor} value is 30, for the magnitude, colour range, and MW components explored in this work.

The \mwf\ code was developed and is currently implemented in the DES-Brazil Portal powered by Laborat\'{o}rio Interinstitucional de e-Astronomia (LIneA\footnote{\url{http://www.linea.gov.br/}}). More details on the DES-Brazil Portal can be found in \cite{2018A&C....25...58G} and \cite{2018A&C....24...52F}. The application of \mwf\  to the DES data took 23h in a SGI ICE-X FC3Y cluster with 4 compute nodes. Each node contained 48 cores and 125 GB of RAM. For more detailed or technical information, the reader is directed to the Appendix B, where the input parameters of the pipeline and details about them are described.

\subsection{Validating the code with mock data}
\label{validation}

\begin{figure}
\centering
\includegraphics[width=230px]{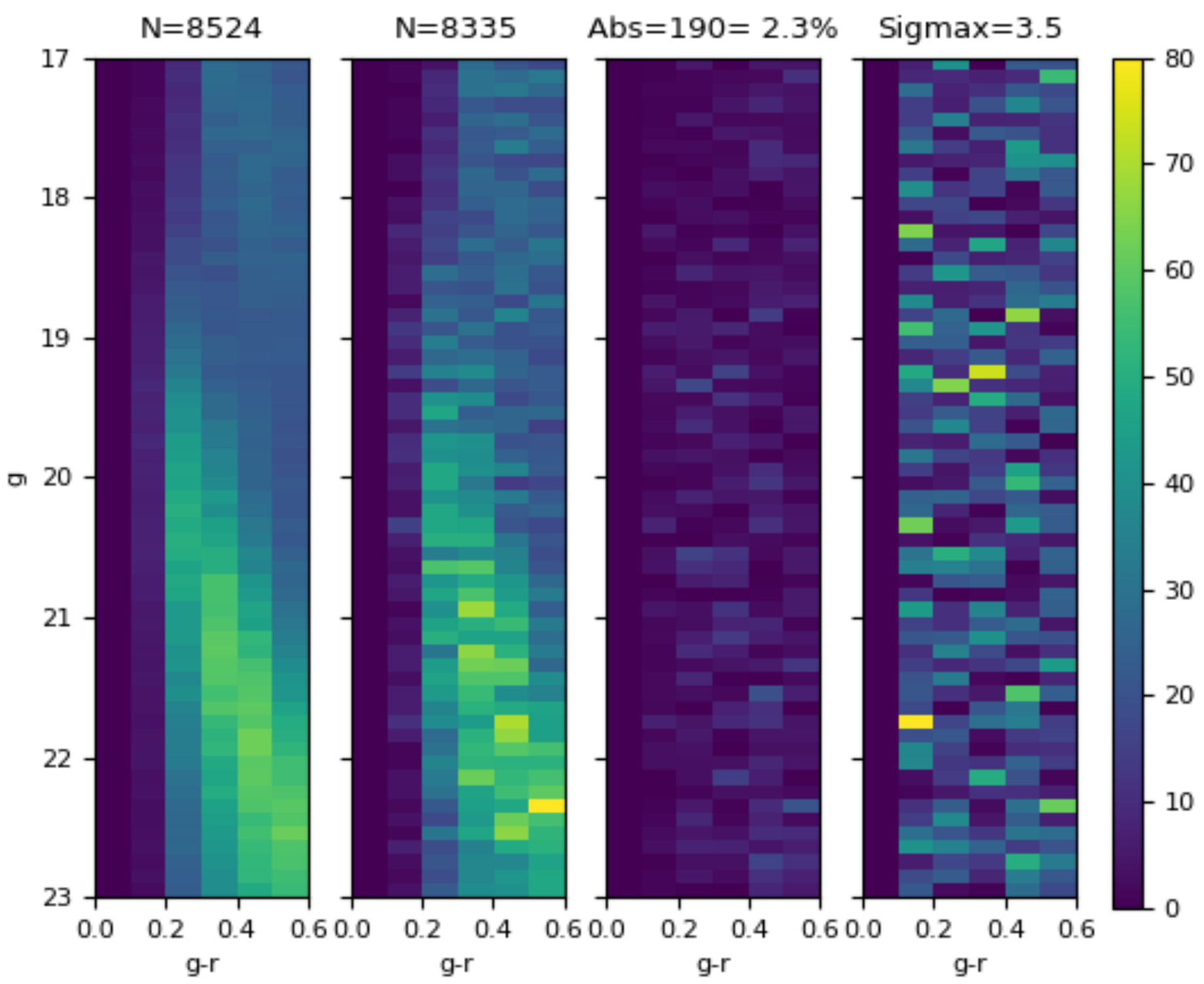}
\caption{HDs for the cell with the largest difference in star counts between the mock data and the best-fit data in test A. \textit{Leftmost panel:} best-fit model. \textit{Second from the left:} input mock data. \textit{Second from the right:} absolute differences between mock data and the best-fit model. These three HDs are colour-coded by star counts according to the colour bar. 
\textit{Rightmost panel:} Poissonian significance, normalized by the maximum significance  ($\sigma_{max}=3.6$). In this case, the colour code is different from that of the colour bar. The title  indicates the number of stars (first and second panel), absolute difference (third panel), and the maximum of the Poisson significance (fourth panel). 
}
\label{simHD}
\end{figure}
\def\arraystretch{2}

In this section we test \mwf\ using mock data. We verify that we can recover the input parameters of our simulated data set when applied to an area with the same footprint as DES-Y3.

Each test utilizes 100 cells, and each cell has the same area as the unit cell designed for the real data (\textsc{HEALPix} pixels with \textsc{nside}=16), following identical footprint and coverage maps (see Section~\ref{sec:data}). The range in magnitude and colour is the same as the DES data ($17<g<23$ and $0.0<g-r<0.6$, respectively), with the same bin size in magnitude and color space (0.1 mag). Uncertainties in the magnitude of the stars were also incorporated in the synthetic data.

Table~\ref{sim} lists the parameters, units, input values, best-fit values and their errors, as indicated by \textsc{emcee}, and the significance of the differences between the best-fit and the true value, for two trials. We run two tests with the same input parameters but different initial values for the MCMC, which we refer to as test A and B. 
Analyzing Table~\ref{sim}, we find that \mwf\ is able to recover the input values of the mock data accurately, even when the initial starting points are far from the true ones. Differences between true and best-fit values are within $3\%$ of the true parameters at the maximum, and the deviations are within 3$\sigma$, with a few exceptions. The maximum differences occur for the density of the halo and its inner exponent, while the differences for the remaining parameters are all below 1\%.

Inspecting the HDs, there is excellent concordance between the mock data and the best-fit model data. The overall range of differences in test A between input data and best-fit models is [-2.28\%, +1.40\%], in terms of star counts. Fig.~\ref{simHD} shows the HDs of the cell with the largest absolute difference (-2.28\%), with the center located at [$l=226.41\degr$, $b=-69.02\degr$]. The panels of Fig. \ref{simHD} shows the HD of the best-fit model, simulated input data (mock), absolute difference, and the Poissonian significance over the HD cells, limited by the maximum significance (given in the title of the panel). The distribution of differences and their significance values show no systematic trend in the colour-magnitude plane. Note that the best-fit HD is smoother than the mock HD distribution due to the oversampling of the model.

Test B produced similar results to test A, with star counts differences in the range [-2.25\%, +2.19\%]. The cell with the largest absolute difference (-2.25\%) exhibits one bin in the HD diagram with maximum significance of 4.6$\sigma$. There is a general concordance in the remaining cells, with typical maximum significance $\leq 4\sigma$ in the cells of the HDs.

The differences between the recovered and true values (the last two columns of Table \ref{sim}) are expected to follow a standard normal distribution, with $\mu$=0 and $\sigma$=1. However, those values appear to be somewhat higher than expected, reflecting a systematic error in the recovery of the true model greater than the uncertainty reported by \textsc{emcee}. In order to encompass half of the recovery errors within $\pm0.67\sigma$ (or 50\% of the area of the standard normal distribution), the uncertainties provided by \textsc{emcee} are increased by a factor of 6.0. In this way, we aim to incorporate realistic systematic errors, and we are assuming they are downsampled by \textsc{emcee} method.

\section{DES Data}
\label{sec:data}

DES  is a wide-area photometric survey covering about $5\,000$ deg$^2$ in the southern Galactic cap \citep{DES2005}. DES images were taken with the Dark Energy Camera \citep[DECam, ][]{2015AJ....150..150F}, with a typical single-exposure (90s in $griz$ bands and 45s in $Y$ band) 10$\sigma$ limiting magnitudes of \emph{g} = 23.57, \emph{r} = 23.34, \emph{i} = 22.78, $z$ = 22.10 and $Y$ = 20.69 for point sources~\citep{2018PASP..130g4501M}. The final coadded images at the end of the first 3 years of observations achieve \emph{g} = 24.33, \emph{r} = 24.08, \emph{i} = 23.44, $z$ = 22.69 and $Y$ = 21.44 at 10$\sigma$ \citep{2018ApJS..239...18A}. DES was designed for cosmological analyses, avoiding the Galactic plane \citep{2018ApJS..239...18A}. Therefore, also considering the depth of the survey, the DES stellar sample will mostly contain stars from the Galactic thick disk and halo. In this section, we characterize the main aspects of the photometry and star/galaxy (S/G) separation in the DES. 

DES-Y3 data were processed by the DES Data Management system \citep[DESDM,][]{2018PASP..130g4501M} and include observations from the first three years of the survey. The DES catalogue studied here is the Year 3 Gold release version 2.2 (Sevilla-Noarbe, in preparation), hereafter referred to as the DES-Y3 catalogue. This catalogue is composed of the same objects as the first public data release \citep[DES-DR1; see][]{2018ApJS..239...18A}, but contains enhanced photometric and morphological measurements and other ancillary information. 

In order to identify the area covered by the DECam observations, the sky is partitioned in \textsc{HEALPix} pixels (\textsc{nside}=4096) with size equal to 52 arcsec $\times$ 52 arcsec (footprint map). Regions around stars brighter than $J=12$ in 2MASS \citep{2006AJ....131.1163S}, globular clusters \citep[][updated 2010]{1996AJ....112.1487H} and a small area close to Large Magellanic Cloud (LMC) were masked. The area covered by DECam in each band and pixel (coverage map) is also estimated by a coverage map produced from \texttt{mangle} \citep{2008MNRAS.387.1391S}. The DES-Y3 catalogue lists objects located in pixels (with \textsc{nside}=4096) with sampled area $>50$\% in \emph{g}, \emph{r}, \emph{i} and \emph{z} bands and imaged at least once in all those four filters.

\begin{figure}
 \includegraphics[width=235px, trim={0.5cm 0.5cm 1.cm 0cm},clip]{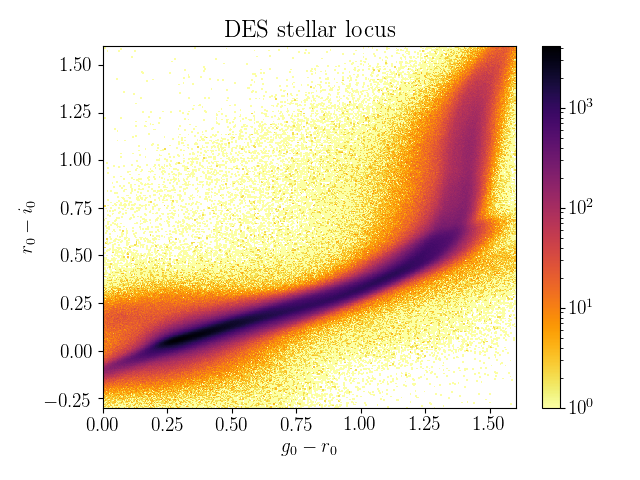}
 \caption{Colour-colour diagram showing the sources selected as stars in DES-Y3 Gold catalogue ($g<23$), following the selection described in the text and corrected for interestellar extinction.}
\label{CCDES}
\end{figure}

The DES-Y3 Gold data is photometrically calibrated by the Forward Global Calibration Method (FGCM\footnote{\url{https://github.com/lsst/fgcmcal}}, see \citealt{2018AJ....155...41B}). A comparison between DES-Y3 and $Gaia$ DR1 \citep{2016A&A...595A...4L} shows a mean difference of 0.0014 magnitudes with $\sigma = 0.0067$ magnitudes~\citep{2018ApJS..239...18A}. The PSF photometry for DES-Y3 catalogue was performed by simultaneously fitting each object in multiple exposures (single object fitting or ‘SOF’). This procedure is very similar to the multi-object PSF-fitting (‘MOF’) described in \cite{2018ApJS..235...33D}. 

Initially, we apply a S/G separation procedure that is  similar to  \cite{2018AAS...23121205S}. 
We use the parameter {\fontfamily{qcr}\selectfont EXTEND\_CLASS\_MASH\_SOF}, which is a variable designed to classify point source (star or quasi-stellar objects - QSO) or extended sources (galaxies) based on {\fontfamily{qcr}\selectfont ngmix} \citep{2015ascl.soft08008S}. We nominally adopt values from the SOF photometry and when SOF photometry is unavailable we adopt photometry from the  coadded images. This criterion increases the stellar sample by including stars with good PSF-fitting in coadded images but with failures in SOF. 
This S/G separation is applied for objects in the full range of magnitudes. Similar to \cite{2018AAS...23121205S}, the same weight-averaged {\fontfamily{qcr}\selectfont SPREAD\_MODEL} in \emph{i} band is applied as S/G classification for the sample of bright stars ($g<18$) where PSF photometry fails.

Extensive completeness assessments were carried out in the DES year 1 (DES-Y1) catalogue, assuring that the catalogue is virtually complete in the range $17<g<22$, with estimated completeness $\geq$ 95\% at the faint limit \citep{2018MNRAS.481.5451S}.

The quality of the DES photometry and S/G classification is illustrated in Fig.~\ref{CCDES}, where we show a colour-colour diagram ($g-r$ vs $r-i$) for sources classified as stars and corrected for reddening following \cite{1998ApJ...500..525S}. There are 13,990,013 sources within the magnitude range $17<g<23$ and the limits shown in Fig.~\ref{CCDES}, namely $0.0 < g_0 - r_0 < 1.6$ and $-0.3 < r_0 - i_0 < 1.6$. A blue plume close to $g_0-r_0 \cong 0$ and $r_0-i_0 \cong 0.25$ amounts to a few thousands stars, probably due to binary systems with a white dwarf and a main sequence star \citep{2004ApJ...607..426K}. A lower level of contamination by QSO's is expected in that region of the color-color diagram ($0.0 \leq g-r \leq 0.5$ and $-0.25 \leq r-i \leq 0.50$), along with contamination on the redder end, which is not taken into account in the process of fitting.

In order to further decrease contamination from mis-classified galaxies, we tested alternative methods for star-galaxy separation. 
The best method that we found was to use the photometric redshift as a way to identify galaxies that were morphologically classified as stars. 
Photometric redshifts were estimated using the Directional Neighbourhood Fitting or DNF \citep{2016MNRAS.459.3078D}, and we refer to this work for details about the fitting of the redshift. We removed objects with DNF photo-z $z>0.55$.

\begin{figure*}
\centering
    \centering
    \begin{minipage}{0.5\textwidth}
    \centering
    \includegraphics[width=250px, trim={3.6cm 1.3cm 2.9cm 0.9cm},clip]{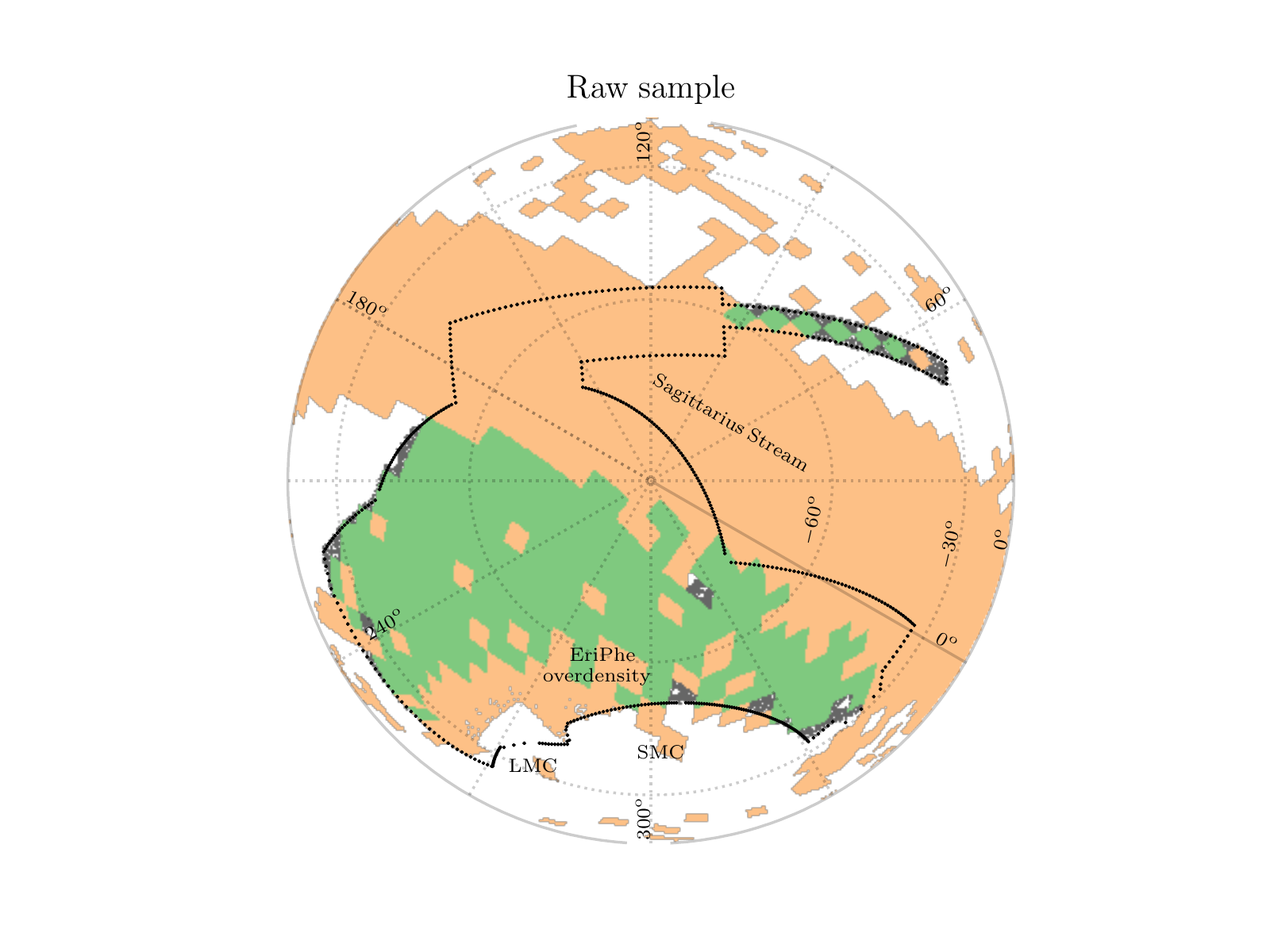}
    \end{minipage}\hfill
    \begin{minipage}{0.5\textwidth}
    \centering
    \includegraphics[width=250px, trim={3.6cm 1.3cm 2.9cm 0.9cm},clip]{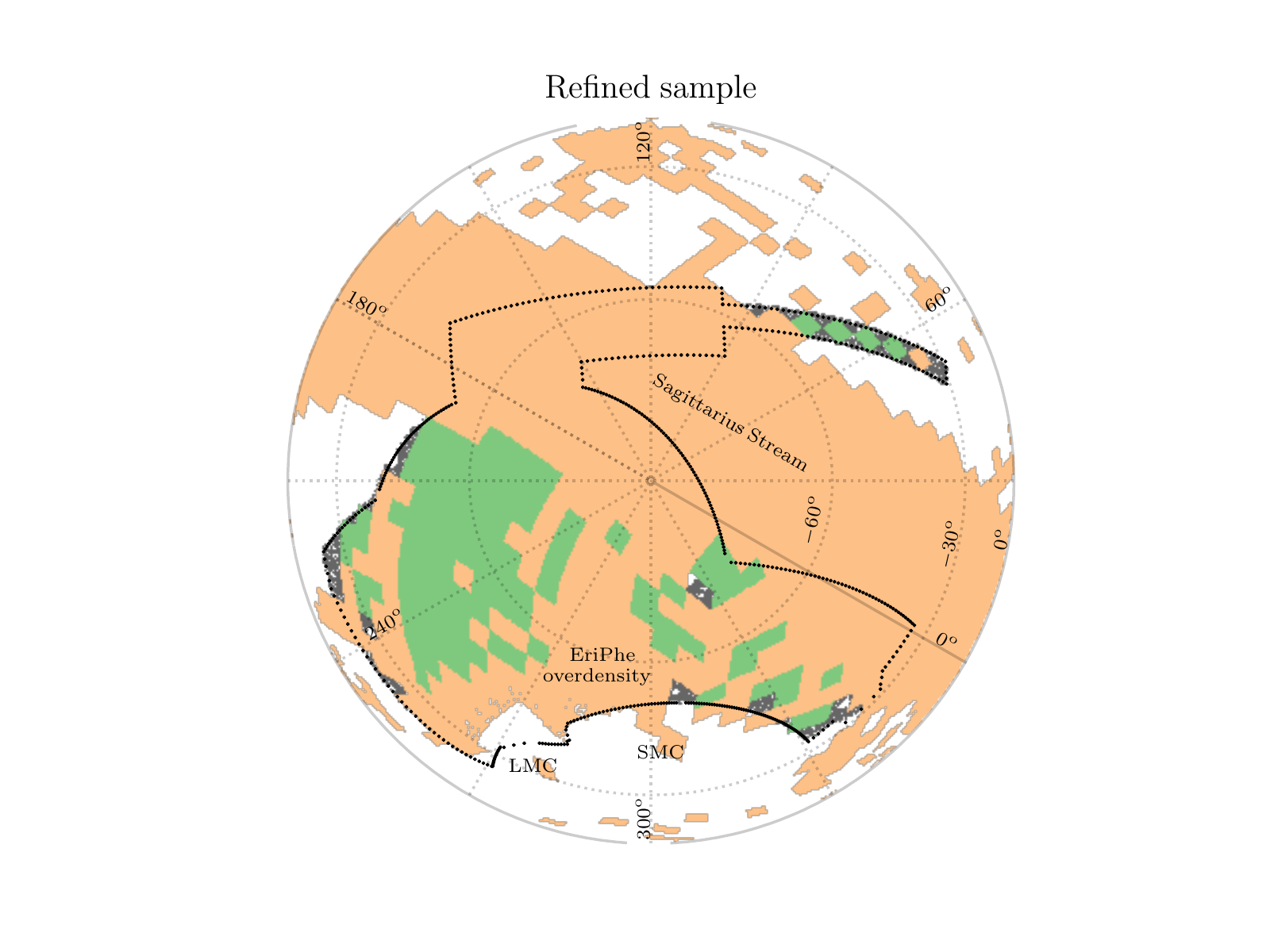}
    \end{minipage}
\caption{Galactic coordinates in an orthonormal projection showing the DES footprint (outlined by black dots) in the southern Galactic Hemisphere. The \emph{raw sample} (\emph{top}) and the \emph{refined sample} (\emph{bottom}) are shown as green diamonds. Cells in orange are masked, due to prominent stellar over-densities such as:  globular clusters, dwarf galaxies, the Sagittarius Stream, the outskirts of the LMC and SMC, Eridanus-Phenix overdensity and stellar streams. LMC and SMC positions are indicated in the figure.}
\label{ftps}
\end{figure*}

To assess the stellar completeness of DES at faint magnitudes, we matched the DES stars to the SPLASH-SXDF catalogue \citep{2019yCat..22350036M}, using as reference for the S/G classification the tag {\fontfamily{qcr}\selectfont STAR\_FLAG}, based on the $BzK$ color-color-diagram. The comparison between DES data and SPLASH-SXDF show DES data is $>$90\% complete down to $g=23$. This confirms the estimates in \citet{2018AAS...23121205S}, and we expect that this sample will have minimal contamination from galaxies and QSOs.

\section{MWFitting applied to DES-Y3 stars}
\label{sec:results}

We partition the DES data into cells corresponding to \textsc{HEALPix} pixels with \textsc{nside}=16, covering a solid angle of $13.43$ deg$^2$. The cells included in the analysis are those with a fill factor $\geq 80$\% ($>10.74$ deg$^2$) of its footprint. Such criterion (and others mentioned below) are identical to those adopted for the validation tests.

We choose a constant range of magnitude ($17 < g < 23$) and color ($0.0 < g - r < 0.6$) when applying \mwf\, to DES data. This constant color-magnitude selection is motivated by the uniformity of the DES footprint in this magnitude depth, and by the high confidence of the modelled stars in this color range. We bin the data in color-magnitude space with a bin size of 0.1 mag in both color and magnitude. This choice of bin size is somewhat arbitrary, and we have found that the results of our analysis are insensitive to the choice of bin size. 

The stars in our sample are not reddening corrected, instead the reddening is incorporated in the models following a Gaussian distribution based on the average and dispersion of the reddening on each cell.

We exclude cells with known stellar clusters and dwarf galaxies.
The list of objects includes globular clusters and dwarf galaxies discovered up-to-date (\citealt[][2010 edition]{1996AJ....112.1487H}; \citealt{2012AJ....144....4M,2015ApJ...813..109D,2015ApJ...805..130K,2015ApJ...808L..39K,2018MNRAS.478.2006L}), along with nearby galaxies partially resolved into stars in the DES images and catalogues (IC5152, ESO294-G010, NGC55, NGC300, NGC1399, NGC247, IC1613, ESO410-G005). The stars from those objects represent a potential contamination to Galactic fields and these fields contained positive residuals in initial iterations  of \mwf .

Cells with any region closer than 22$^{\circ}$ from the LMC centre were also masked. \cite{2019ApJ...874..118N} clearly shows (see their figure 5) a significant population of LMC main-sequence stars located out to 21$^{\circ}$ from the center of the LMC. Furthermore, we masked the Sagittarius Stream, removing a stripe of width equal to 20$^{\circ}$ along the centre of the stream \citep{2003ApJ...599.1082M}.

After removing the aforementioned regions and selecting only cells with a fill factor of more than 80\%, the remaining 194 cells constitute our so-called \emph{raw sample}. This sample includes the stellar population of streams discovered in the DES footprint \citep{2018AAS...23121205S} and the Eridanus-Phenix overdensity \citep[Eri-Phe, ][]{2016ApJ...817..135L}. Since these objects cover a large area with a much lower stellar density than that of the Galaxy, we retain them in the \emph{raw sample}. 

\subsection{With or without streams?}

To explore the influence of including regions with known stellar streams and the Eri-Phe overdensity, we define a second sample removing the regions where those objects are located. The list of masked stellar streams is that described by \cite{2017ascl.soft11010M}, and we refer to this work for further details. In the case of Eri-Phe over-density, the masked area has a triangular shape as shown in figure 3 of the discovery paper \citep{2016ApJ...817..135L}. The second sample of DES data comprise 105 cells, and we refer to this sample as the \emph{refined sample}.

Fig.~\ref{ftps} puts into perspective the footprint of the \emph{raw} and \emph{refined} samples using an orthonormal projection of the southern Galactic Hemisphere. 
The DES footprint is outlined in black. 
The cells included in \mwf\ are displayed in green and masked cells are shown in orange.  The \emph{raw} and \emph{refined} samples are top and bottom respectively.  A significant portion of the DES footprint is masked in the  \emph{refined} sample. 

The Sagittarius Stream stands out in both panels of Fig.~\ref{ftps} as a wide stripe crossing the South Galactic Pole and cells masked due to proximity to the LMC are in the lower left corner. 
The area sampled by DES-Y3 and compared to models amounts to 2,315 deg$^2$ (194 cells) in the \emph{raw} sample, and to 1,256 deg$^2$ (105 cells) in the \emph{refined} sample.

\begin{table*}
\begin{center}
\setlength{\tabcolsep}{6pt}
\caption{Best-fit parameters for the \emph{raw} and \emph{refined} samples. The last two columns are results from the literature. In our results, the first errors listed are the $1\sigma$ statistical error or the standard deviation of the mean estimated by the \emph{jackknife block} method (see more details in the text). The second errors are the systematic errors as discussed in Section~\ref{sec:sim} and ~\ref{res}. They represent the ability of the pipeline to recover the true model, and the degeneracy of the parameters regarding the uncertainty of the local density of the thin disk.}
\begin{tabular}{l l r r r r r r}
\hline
\multirow{2}{*}{\textit{Parameter}} & \multirow{2}{*}{\textit{Unit}} & \multicolumn{2}{c}{\mwf\ } & 
\multicolumn{1}{c}{Juri$\mathrm{\acute{c}}$} & \multicolumn{1}{c}{de Jong} & \multicolumn{1}{c}{Deason}\\
 &  & \textit{Raw sample} & \textit{Refined sample} & et al. 2008 & et al. 2010 & et al. 2011 \\
\hline
ThickDisk $h_e$ & pc & ${925}\pm{6}\pm{40}$ & ${910}\pm{8}\pm{45}$ & $743\pm150$ & $750\pm70$ & -\\ 
ThickDisk $R_e$ & pc & ${2667}\pm{89}\pm{34}$ & ${2631}\pm{111}\pm{49}$  & $3261\pm650$ & $4100\pm400$  & - \\ 
ThickDisk $\rho$ (R=R$_{0}$) & $10^{-3}$ $\Msun$pc$^{-2}$ & ${3.89}\pm{0.09}\pm{0.64}$ & ${3.97}\pm{0.12}\pm{0.73}$ & $7.53\pm0.75$ & $5.01\pm1.30$  & - \\
Halo $n_1$ & - & ${1.82}\pm{0.05}\pm{0.06}$ & ${1.86}\pm{0.07}\pm{0.08}$ & - & - & $2.3^{+0.1}_{-0.1}$ \\
Halo $n_2$ & - & ${4.14}\pm{0.03}\pm{0.04}$ & ${4.24}\pm{0.04}\pm{0.05}$ & - & - & $4.6^{+0.2}_{-0.1}$ \\
Halo $r_{br}$ & kpc & ${18.52}\pm{0.15}\pm{0.23}$ & ${18.59}\pm{0.39}\pm{0.29}$ & - & - & $27^{+1}_{-1}$ \\
Halo $q$ & - & ${0.75}\pm{0.01}\pm{0.01}$ & ${0.74}\pm{0.02}\pm{0.01}$ & $0.64\pm0.01$ & $0.88\pm0.03$ & $0.59^{+0.02}_{-0.03}$ \\
Halo $\rho$ (R=R$_0$) & $10^{-5}\,\Msun\,pc^{-3}$ & ${3.31}\pm{0.10}\pm{0.17}$ & ${3.51}\pm{0.13}\pm{0.23}$ & ${2.95}\pm{0.74}$ & ${6.31}\pm{0.77}$ & - \\ 
\hline \\
\end{tabular}
\label{finaltable}
\end{center}
\end{table*}

\subsection{MWFitting configuration and errors}

Before discussing the outcomes from applying \mwf\ to DES data, we first discuss the \textsc{emcee} configuration used.
We use 200 walkers along 250 steps with step length as 1\% of each parameter to sample the posterior distribution. 
We perform initial iteration, starting with input values from the literature.  In a second iteration, we redo the fit starting from previous fitting. The first 200 steps are discarded as a burn-in phase, and we examine the remaining distribution to check that the walkers have converged. We apply a Gelman-Rubin convergence diagnostic ($R_c \leq 1.004$) to check for convergence of the Markov chains.

The results from applying \mwf\ to the \emph{raw} and \emph{refined} samples are listed in Table~\ref{finaltable}. We find that the errors reported from the posterior distribution are smaller than the difference of best-fit parameters when we tested the pipeline with subsets of the \emph{raw} or \emph{refined} sample. 

Hence, we have decide to estimate the statistical errors from a \emph{jackknife} resampling method \citep{Feigelson2012}, in addition to the systematic errors based on the \textsc{emcee} method. The \emph{jackknife} method creates $n$ samples (where $n$ is the number of observations), replicating the initial sample in each iteration, but omitting the i-$th$ observation. The \emph{jackknife block} method is  similar, but instead we group the observations into $n_b$ data blocks with size $k$ (in our case, the blocks are a set of cells). In each $i$ subsample with $k$ size, a pseudo-value $ps_i$ is calculated: 

\begin{equation}
ps_i(X) = n_b \phi_n(X_1,...,X_n)-(n_b-1)\phi_{n-k}((X_1,...,X_n)_{[i]})
\end{equation}
\noindent
where $\phi_n$ is the statistical estimator (e. g. mean or dispersion) defined for $n$ blocks and $\phi_{n-k}((X_1,...,X_n)_{[i]}$ is the same estimator but for the \emph{deleted-one} sample. The pseudo-values, $ps_i$, follow a standard normal distribution for the $\phi$ parameter with mean and standard deviation. 

We adopted $k=10$ for both samples, with $n_b=20$ blocks in the \emph{raw} and $n_b=10$ blocks in \emph{refined} sample. Following this method, the statistical errors indicated in Table~\ref{finaltable} bound 1$\sigma$ or 68\% of the likelihood distribution of each parameter. 
One potential concern is that imperfect modelling of the thin disk could affect fitted parameters of the thick disk and halo. In order to assess this possible degeneracy, we run multiple fits of the halo and thick disk with the thin disk density set to 60\%--110\% (with bin size equal to 10\%) of the benchmark value listed in Table \ref{tabparameters} (55.41 M$_{\odot}$ pc$^{-2}$). Assuming an uncertainty of 10\% in the local surface density of the thin disk (similar to the uncertainty of \citealt{2004MNRAS.352..440H}), those trials indicate a strong dependence between the densities of thin/thick disk. A decrease of 10\% in the modelled density of the thin disk means an increase of the same amount in the density of the thick disk, while for the remaining parameters the dependence is much weaker. In this way, we assume an uncertainty of 10\% in the local density of the thin disk, and we added the systematic dependence on the thin disk local density as an systematic error in Table \ref{finaltable} for all the parameters. We assume that the correlation between the parameters in Table \ref{finaltable} and the parameters of the thin disk (with the exception of the density of the thick disk) is much weaker than the correlation between the same parameters and the normalization of the thin disk.

Following this reasoning, the systematic errors included in Table \ref{finaltable} account for the ability of the pipeline to recover input values, and the dependence of the parameters on the local density of the thin disk.
The best-fit parameters for the \emph{raw} and \emph{refined} samples agree within 1$\sigma$ and have quite similar errors.

\subsection{MWFitting results}
\label{res}



There is a general agreement between our results and previous works (see Table~\ref{finaltable}), even though our uncertainties are smaller in most of the cases.

The vertical and radial scale of the thick disk are consistent within $\sim 1\sigma$ given the estimate and uncertainty from~\cite{2008ApJ...673..864J}, and the density normalization of the thick disk is within 1$\sigma$ of the estimate and uncertainty from~\cite{2010ApJ...714..663D}.

The large differences in the density of the thick disk reported by previous works may be related to the different methods used to estimate the total stellar mass. Different IMFs heavily influence the number of low-mass stars, most of which are not sampled by the HDs in this work. Different approaches in selecting stars also impact the estimation of the total stellar mass. Likewise, we point out there is a discrepancy by a factor of $\sim$2 in the local halo stellar density between the estimations of \cite{2008ApJ...673..864J} and \cite{2010ApJ...714..663D}.

Comparing our measurements of the Galactic halo to the literature, the best-fit values of oblateness ($q$) are between the results of \cite{2008ApJ...673..864J} and \cite{2011MNRAS.416.2903D} and that of \cite{2010ApJ...714..663D}. 
Regarding the inner and outer exponents of the double power law describing the halo density, we find that estimates from \cite{2011MNRAS.416.2903D} are steeper than ours, but that the two results are consistent to within 20\%. This relative discrepancy could be due to many factors: minor tweaks in the stellar evolutionary models, the different regions sampled (SDSS imaged most the northern hemisphere while DES samples the southern hemisphere), or the other model parameters adopted. Similar explanations could account for the difference between our results and the single power-law fit by \citealt{2018ApJ...859...31H} (n=$4.40^{+0.05}_{-0.04}$), in addition to the fact that they use RR Lyrae stars from Pan-STARRS1 between 20 kpc $\leq R_{GC} \leq$ 131 kpc, which extend to much larger distances than our sample ($r_{GC} \leq 60$ kpc).

\begin{figure*}
\centering
 \includegraphics[width=460px]{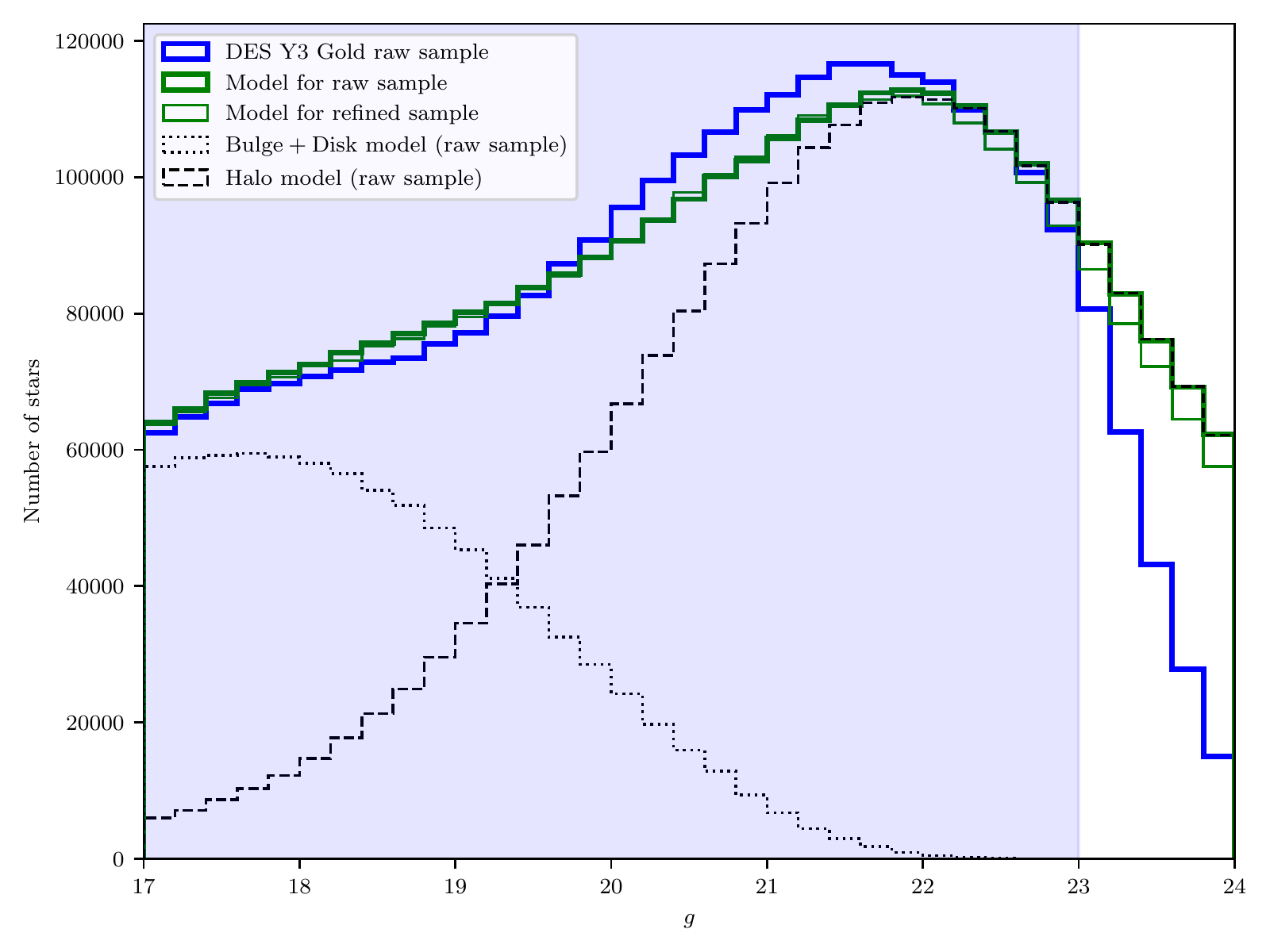}
 \caption{Stellar number distribution in \emph{g}-band for the DES-Y3 catalogue (blue line) and four different models (green and gray histograms). 
 The best-fit model for the \emph{raw} and \emph{refined} samples are shown as thick and thin green lines, respectively. In grey, we show the same model as the raw sample, but splitted in two main components: disk (dotted line, with a small contribution of the bulge) and halo (dashed line).}
\label{histmagg}
\end{figure*}

Our model indicates a closer power-law break radius than that indicated by \cite{2011MNRAS.416.2903D}; however, our best-fit break radius is consistent with the larger range of fits in the literature. In order to illustrate the range of distances for the radius of the break in previous works, we cite a few examples using diverse methods. For example, \cite{2009MNRAS.398.1757W} use a sample of RR Lyrae stars in Stripe-82 region sampled by SDSS, finding a break radius of 23 kpc. \cite{2015A&A...579A..38P} fit F stars from fields of MENeaCS and CCCP projects determining a power-law break at 20 kpc from the Galactic center, and in a more recent work \cite{2015ApJ...809...144} modelled giant stars from SDSS/SEGUE-2 found a closer break radius than our value ($18\pm1$ kpc).

In a more recent work, \citet{2018ApJ...862L...1D} determined the orbital properties of a sample of MS and BHB halo stars using position, kinematic properties and metalicites from Gaia DR2 and SDSS. Adding the Galactic gravitational potential, they derive the apocenter of the star's orbits, addressing the break of the halo to a ``pile-up'' effect where the stars with eccentricity $e > 0.9$ slow-down near the most distant part of the orbit. After excluding stars from the disk, the average apocenter derived for MS and BHB stars are 16$\pm$6 kpc and 20$\pm$7 kpc, respectively, in excellent agreement with our fit (see also~\citealt[][]{2005ApJ...635..931B, 2013ApJ...763..113D} and references therein).

We also fit two alternative models for the Galactic halo: an Einasto profile and a single power law (in both cases the thick disk was modeled with the same exponential profile). The best-fit model with halo modeled by an Einasto profile yielded a lower likelihood ($-2\ln \lambda = 440{,}340$) than the model with the double power law ($-2\ln \lambda = 218{,}355$), both following Eq.~\ref{like}. The best-fitting model for thick disk and halo with a single power law resulted in an even lower likelihood ($-2\ln \lambda \simeq 1{,}000{,}000$). These conclusions are quite similar to those of \cite{2011MNRAS.416.2903D}.


N-body simulations \cite[e.g.,][]{2005ApJ...635..931B} show that an excess of stars in the central region of the halo (similar to the observed double power law of the halo) can be related to accretion events. Thus, the features of the halo's stellar profile provide clues about the epoch, number an characteristics of past accretion events. Following this reasoning, the observed features of the best-fit halo model strongly favour a massive accretion event where the stars from the accreted satellite dominates the Galactic halo out to the break radius. We posit that these stars may be associated with the merger of the Gaia-Enceladus-Sausage galaxy \citep{2018MNRAS.478..611B, 2018Natur.563...85H}. The stellar mass derived from integrating the best-fit model for the $raw\ sample$ from the Galactic center out to the break radius results in a mass of $\simeq 3.6 \times 10^{8}\,M_{\odot}$. This estimate can be used as an upper limit for the current stellar mass of the Gaia-Enceladus-Sausage galaxy, excluding possible globular clusters associated to the former satellite. In a comparison to a recent work, \cite{2020MNRAS.492.3631M} selected a mono-abundance population of halo ($-3<$[Fe/H]$<-1$ and $0.0<$[Mg/Fe]$<0.4$) red giant stars from APOGEE-DR14, and they found a estimation of the current mass of stars with high eccentricity ($e > 0.7$) associated to Gaia-Enceladus-Sausage galaxy ($3\pm1\times 10^{8}\,M_{\odot}$) very close to our estimate above.

\section{Simulating the stellar contents of DES-Y3}
\label{sec:sim}

\begin{figure*}
\centering
 \includegraphics[width=0.99\textwidth, trim={1cm 0.2cm 0cm 0cm},clip]{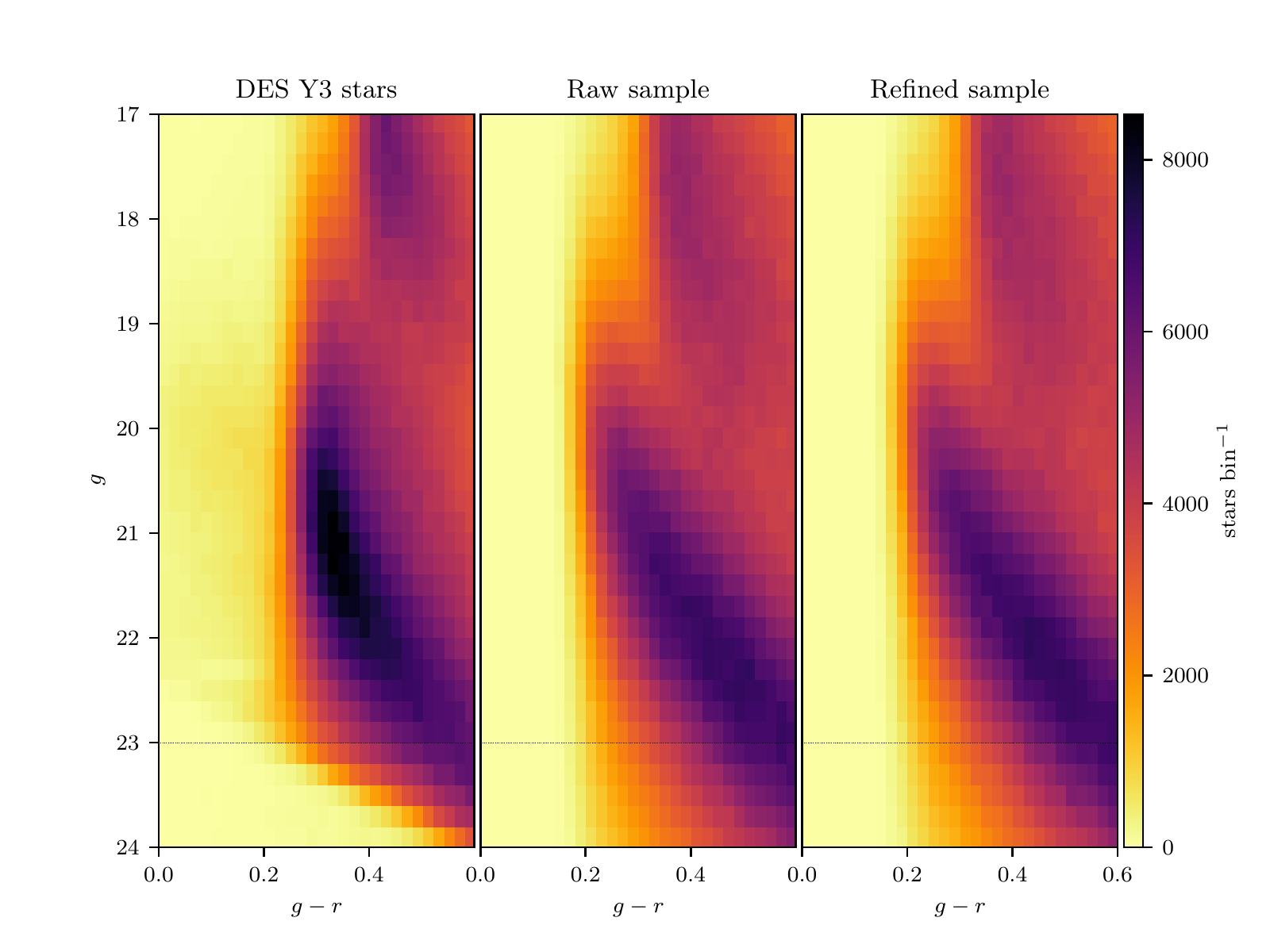}
 \caption{\emph{Left panel:} CMD for the \emph{raw} sample of DES-Y3 stars (blue line in Fig.~\ref{histmagg}). \emph{Central panel:} simulated CMD for the \emph{raw} sample. \emph{Right panel:} simulated CMD for the refined sample. The cut in photometric redshift explained in Sec.~\ref{sec:data} is responsible for the reduced source density at the faint end of the left panel.}
\label{CMDDR1break}
\end{figure*}

With the best-fit parameters, we produce a simulated stellar catalogue matched to DES-Y3 with limiting magnitude of $g=24$ and in the colour range $0<g-r<0.6$. We compare these simulations to the real data to study the stellar distribution in DES-Y3, to highlight asymmetries in the Galactic components (such as flares and warps in the disk), and to reveal stellar substructures.

Fig.~\ref{histmagg} compares the star counts as a function of \emph{g} magnitude in DES-Y3 to simulations using the best-fit models. The regions where the stars are sampled exclude areas containing dwarf galaxies, globular clusters, stellar streams, the Sagittarius Stream and Eridanus-Phoenix over-density, and regions with high reddening ($b \leq -30^{\circ}$). The magnitude bins in this figure are twice the size of the magnitude bins in the fitting, in order to sample a smooth histogram.

The distribution of DES-Y3 stars in Fig.~\ref{histmagg} is shown as a blue line, while the distribution of stars in the simulations using the best-fit parameters from the \emph{raw} and \emph{refined} samples are shown as thick and thin green lines, respectively. Grey lines sample the distribution of modelled stars belonging to the bulge and disk (dotted line) or to the halo (dashed line), both following the best-fitting model for \emph{raw} sample.


An initial look at Fig.~\ref{histmagg} reveals a high level of consistency between the two best-fit models. The differences between both models are $<\ 2$\% in general. These models are reasonably similar to the data, agreeing within 5\% in the magnitude range $17<g<23$. The histogram shows an excess in the DES-Y3 data with respect to both best-fit models between $20<g<22$, with an excess in the modelled stars of a few percent between $18<g<19.5$. The discrepancies between data and model in Fig.~\ref{histmagg} may be improved in several ways: 
better models for the evolution of metal-poor stars (population of the halo), minor tweaks in the halo's SFH, additional components in the Galactic halo model (e.g., the Gaia-Enceladus galaxy), a potential metalicity gradient in the halo, slight changes to the Sun-Galactic center distance, or in any other parameter taken into account in the \trilegal  models, such as the interstellar extinction. We are investigating the possible causes for the observed excess of stars, in order to develop an improved model for the Galaxy. Interestingly, the difference between data and models is dependent on the region of the sky examined, with better agreement found including only Galactic fields at higher latitude.

\begin{figure*}
\centering
 \centering
 \begin{minipage}{0.5\textwidth}
 \centering
 \includegraphics[width=290px, trim={5.6cm 2.5cm 4cm 0.5cm},clip]{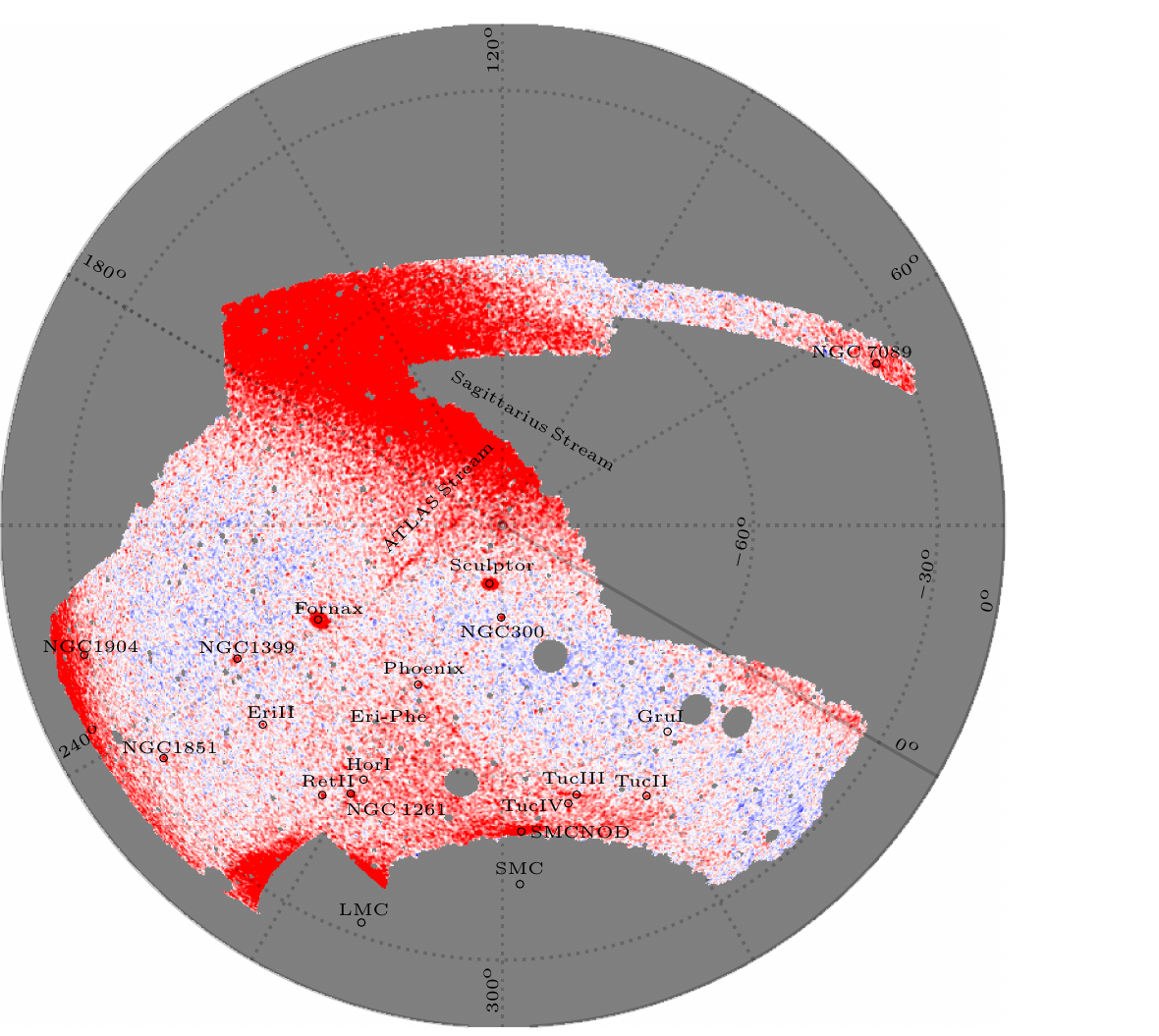}
 \end{minipage}\hfill
 \begin{minipage}{0.5\textwidth}
 \centering
 \includegraphics[width=290px, trim={5.6cm 2.5cm 4cm 0.5cm},clip]{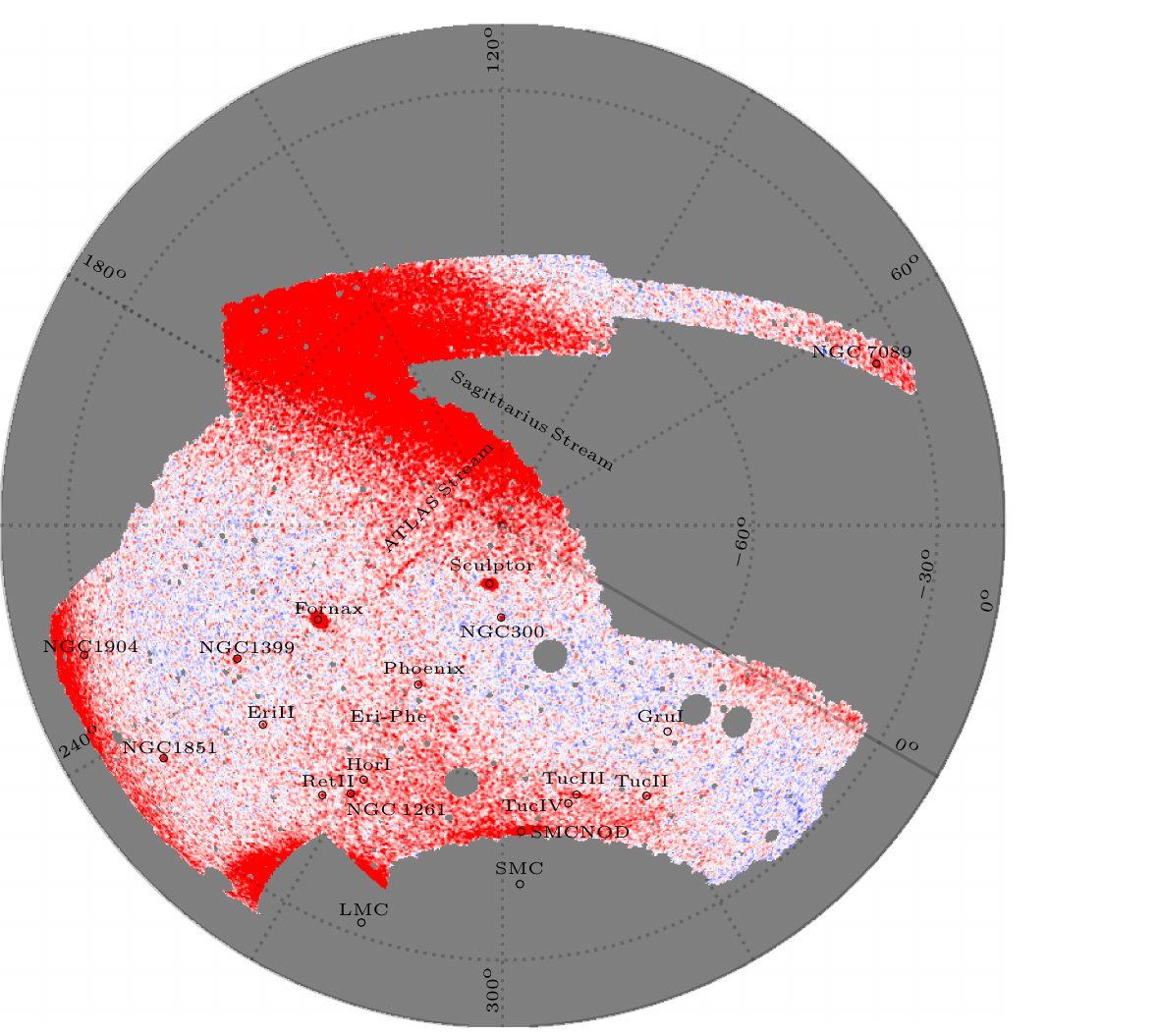}
 \end{minipage}
 
 \begin{minipage}{1.0\textwidth}
 \centering
 \includegraphics[width=400px, trim={4.cm 1.cm 4cm 11.8cm},clip]{_GAL_orth_194f.pdf}
 \end{minipage}
 
\caption{Smoothed Poisson significance ([$N_{obs}-N_{mod}]/\sqrt{N_{mod}}$) of residual maps between the DES-Y3 stars and best fit MW models created with the \emph{raw} (left) and  \emph{refined} (right) samples, with a limiting magnitude of  $g=23.5$. The significance value in each cell is smoothed using a Gaussian kernel with full-width-half-maxima $\simeq$ 7 arcmin. 
Many over-densities are identified, most of them are associated with known objects including globular clusters, dwarf galaxies and stellar streams. The insertion of more labels in the figure were avoided to do not pollute excessively the map.
Both maps are set to the same scale and regions masked (not covered by DES or close to bright stars) are shown in gray. Despite the fact that we are not fitting the thin disk and the bulge, the overall smoothed Poissonian significance across the footprint is close to zero. 
}
\label{DESEQU}
\end{figure*}

Fig.~\ref{CMDDR1break} shows the same distribution of stars, but in the $g \times g-r$ CMD space, with bins in magnitude and color equal to 0.2 and 0.02, respectively. We note that this bin size is different from that used in the fitting, but is equal to that in Fig.~\ref{histmagg} for the $g$ magnitude. In order to highlight subtle differences in color, we oversample the $g-r$ color range with bin sizes equal to 0.02.

The left panel of Fig.~\ref{CMDDR1break} shows the distribution of DES-Y3 stars, similar to the blue line in Fig.~\ref{histmagg}. The best-fit model for the \emph{raw} and \emph{refined} samples are shown in the central and right panels representing the green lines in Fig.~\ref{histmagg}. Analogous to the observed magnitude distribution at the faint end in Fig.~\ref{histmagg}, the fainter end of the first panel in Fig.~\ref{CMDDR1break} shows a decreasing number of sources below $g=23$, where the dashed line delimits the bound of the stars compared to models in the fitting. The panels share the same colorbar, indicated on the right of the figure.

The three panels of Fig.~\ref{CMDDR1break} exhibit strong similarities down to $g \lesssim 23$. The thick disk leaves its main imprint by the plume of MSTO metal-rich stars at $g < 19$ and $g-r \simeq 0.4$. There is a smooth transition between the crowding of MSTO stars of the thick disk and the MSTO stars of the halo, which starts at $g \simeq 19$ but in a bluer region. This transition is seen in the Fig.~\ref{histmagg} as a distribution of stars slightly more flat ($18 < g < 19$) than the preceding or subsequent range. The MSTO halo's stars are concentrated in a large range of magnitudes centered at $g \simeq 21$, whose density smoothly decreases towards the fainter end.

An excess of DES-Y3 stars in the range $21<g<22$ is seen on the left panel of Fig.~\ref{CMDDR1break}, similar to Fig.~\ref{histmagg}, but with the additional information that the excess stars are concentrated near the MSTO of halo stars. The most populated bin in the left panel of Fig.~\ref{CMDDR1break} ($g-r$=0.33, $g=21.25$) contains 25\% more stars when compared to the same bin in the central panel.

Even though they are not included in this comparison, the estimation of star counts fainter than $g=23$ is important for future surveys such as the Rubin Observatory LSST~\citep{2009arXiv0912.0201L} and Euclid~\citep{2016MNRAS.459.1764S}, where S/G classification will be important.
For example, at $g=24$, Fig.~\ref{histmagg} shows that the expected number of halo stars in the models is roughly double the number of stars in the data. 
Realistic simulations for future large and deep surveys must consider and account for this incompleteness.

\subsection{Poissonian significance maps}

Fig.~\ref{DESEQU} shows the Poissonan significance maps generated for both samples of the DES-Y3 data using the best-fit model parameters. 
Given the steep decrease of stars at faint \emph{g}-magnitudes in Fig.~\ref{histmagg}, we restrict the sample to stars with $g<23.5$.
The significance of each $7\times7$ arcmin$^2$ pixel is taken as the residual star counts (difference between the DES-Y3 catalogue and the model catalogue) divided by the square root of modelled star counts. Both maps are smoothed by a Gaussian kernel with $\sigma$ = 7 arcmin, resulting in a minimum significance of -1.67 for \emph{refined sample} and -1.69 for \emph{raw sample}. In order to highlight under/overdensities as blue/reddish colours, and white colour as a perfect agreement between models and data, the significance range is set to [-1.67,1.67]. Pixels with significance higher than 1.67 (mainly known globular clusters and dwarf galaxies) are saturated with that maximum value.

Many known Galactic substructures are enhanced in this residual map, attesting to the accuracy of the \mwf\ model. 
We label the most significant stellar over-densities on both panels of Fig.~\ref{DESEQU}. For instance, the stripe roughly parallel to $l=180^\circ$ is the Sagittarius Stream. The over-density associated with SMC (SMCNOD) in the anti-LMC side \citep{2017MNRAS.468.1349P,2018ApJ...858L..21M} is also evident.
Although we are not using a $matched\ filter$, a technique commonly applied to highlight fainter substructures as streams \citep[e.g.,][]{2003AJ....126.2385O}, a few streams are noticeable in Fig.~\ref{DESEQU}. The ATLAS stream \citep{2014MNRAS.442L..85K,2018AAS...23121205S}, a track of stars close to Galactic Pole (indicated in Fig.~\ref{DESEQU}), is a good example of such a structure, as well as the Phoenix stream \citep{2016ApJ...820...58B}, a long track of stars seemingly pointing toward the Phoenix dwarf galaxy. Other visible features are the Indus stellar stream (just below Tuc II), and the Tuc III stream, centered on the dwarf galaxy Tuc III.

The regions at the lowest Galactic latitudes between $240^\circ<l<270^\circ$ presents smooth and flat over-densities (with the exception of the region close to LMC) in both panels of Fig.~\ref{DESEQU}, which may indicate that there is room for improvement in the thin disk model. The region at $b<-30^\circ$, $220^\circ<l<240^\circ$ in DES-Y3 footprint exhibits a strong excess of stars close to NGC1904, which may be the result of disk flaring or the Southern extension of the Monoceros Ring \citep{2002ApJ...569..245N}.

The Eridanus-Phoenix over-density \citep{2016ApJ...817..135L} is a very large over-density of stars between $270^\circ<l<330^\circ$ and $-40^\circ<b<-70^\circ$, populating a triangle with vertices close of LMC, SMC and Fornax dwarf galaxy, seen on both panels of Fig.~\ref{DESEQU}. Subtracting the stars in the modelled catalog, the Eridanus-Phoenix cloud contains an over-density of 4756 (4755) stars within the range ($17<g<22$ and $0.0<g-r<0.6$) when compared to the best-fit of the \emph{raw} (\emph{refined}) sample. Accounting for stars more massive than 0.1 M$_{\odot}$ in a Chabrier mass function \citep{2000ApJ...542..464C} for a disk-like IMF stars, those values correspond to an object with $\simeq 1.6 \times 10^4$ ($\simeq 1.5 \times 10^4$) M$_{\odot}$ for the \emph{raw} (\emph{refined}) sample. These mass estimations represent a decrease in mass of at least by factor of five compared to the estimates in \citet{2016ApJ...817..135L}.

Even though the best-fit parameters for both samples agree within 1$\sigma$, there are slight differences regarding the two panels of Fig~\ref{DESEQU}. For instance, ATLAS and Phoenix streams seem to be more continuous in the left panel with best-fitting parameters from \emph{raw} sample than with the \emph{refined} sample.


\subsection{Milky Way stellar mass}

We calculate the stellar masses of the halo and thick disk MW components with the best-fit parameters (Table~\ref{finaltable}) and list them in Table~\ref{MWmass}. These mass estimations only include field stars following from a smooth model for the Galactic components, and therefore exclude  the mass from globular clusters, dwarf galaxies, and streams. 

The bulge parameters are kept fixed, and the model described in Table~\ref{tabparameters} amounts to a stellar mass of $1.28 \times 10^{10}$ M$_{\odot}$ or 21.4\% of the total stellar mass of the Galaxy ($5.97\pm0.99 \times 10^{10}$ M$_{\odot}$, following the adopted model here). 
This agrees with mass estimates from the literature, where estimates of the stellar bulge mass range from 10-20\% of the MW stellar mass~\citep{2015ApJ...806...96L,2017MNRAS.465.1621P}. 
Our model includes a thin disk (with fixed parameters) and has a stellar mass of $4.62 \times 10^{10}$ M$_{\odot}$, which is within 1$\sigma$ of the estimation by \citealt{2015ApJ...806...96L} ($5.17 \pm 1.11 \times 10^{10}$ M$_{\odot}$) and within 2$\sigma$ of \citealt{2011MNRAS.414.2446M} ($5.54 \pm 0.63 \times 10^{10}$ M$_{\odot}$). The thick disk has a small contribution to total disk mass, with a ratio of stellar masses in the thin and thick disks as $\cong 13\,000:1$. 

The halo mass is estimated by integrating the double power-law profile from the Galactic Centre out to 100 kpc. Based on our fits, we find that the stellar halo contributes 1.1\% of the Galactic stellar mass, while the disks contribute with $\simeq$80\% of the total. Our estimate of the total stellar halo mass is within the range estimated by \citealt{2011MNRAS.416.2903D} ($2-10 \times 10^8$ M$_{\odot}$), while being more massive than estimated by \citealt{2008ApJ...680..295B}, where the latter authors found a halo with an integrated stellar mass out to 40 kpc of $3.7\pm1.2 \times 10^8$ M$_{\odot}$.

\begin{table}
\begin{center}
\setlength{\tabcolsep}{6pt}
\caption{Stellar masses estimates for the MW components fit in this work, for the \emph{raw} and \emph{refined} samples.}
\begin{tabular}{l c c}
\hline
\multirow{2}{*}{Component} & \multicolumn{2}{c}{Estimated mass (M$_{\odot}$)} \\
 & \emph{Raw sample} & \emph{Refined sample} \\
\hline
Thick disk & $3.57 \pm 0.43 \times 10^{6}$ & $3.70 \pm 0.44 \times 10^{6}$ \\
Halo (r$<$100kpc) & $6.80 \pm 1.04 \times 10^{8}$ & $6.98 \pm 1.56 \times 10^{8}$ \\
\hline
\end{tabular}
\label{MWmass}
\end{center}
\end{table}




    \section{Concluding Remarks}
    \label{sec:conc}

We have developed a new code to fit the stellar components of the MW. In this first paper, we concentrate on fitting the thick disk and the halo due to the location of the DES footprint in the south Galactic cap. We list below our main conclusions from this work.

\begin{itemize}
\item
This work presents \mwf, a pipeline constructed to fit structural parameters for the Galactic components with \trilegal\ stellar population synthesis models.
\item
The \mwf\, pipeline is validated with synthetic catalogues. We successfully recovered the input parameters (with a maximum deviation $\leq$ 3\%) using the same oversampling factor and a footprint smaller than the real data (see Table~\ref{sim}).
\item 
Our main goal in this work is to model the halo and the thick disk components by applying the \mwf\ pipeline to data from DES-Y3 Gold catalogue. We defined two different samples based on known stellar over-densities. Both samples excluded cells populated by dwarf galaxies, globular clusters and cells close to the LMC. In the \emph{refined} sample, we further excluded cells where stellar streams and Eridanus-Phoenix over-density are located.
\item
Table~\ref{finaltable} lists the results for both samples, with statistical uncertainties determined by \emph{jackknife} resampling and the \textsc{emcee} method. The systematic uncertainties are sampled by the ability of the pipeline in recovering the true parameters based on simulations and the uncertainties in the local density of the thin disk. Results from both samples agree within a confidence level of 68\% (1$\sigma$).
\item
The distribution of DES-Y3 stars presents a reasonable agreement (within $\leq 5\%$ in number of stars in each bin) with our models down to $g=23$. The distribution of stars in the DES-Y3 catalogue and in the models both peak close to $g=22.0$. Fainter than $g=23$, there is a decrease in the number of stars, that we interpret as a result of the S/G classification schema applied here, coupled with the relative scarcity of stars in the outer MW halo.
\item
CMDs comparing DES-Y3 stars and both simulations reasonably agree down to $g = 23$, suggesting that the double power law is a good description of the Galactic halo, at least at this depth.
\item
The star counts in the stellar halo is crucial for predicting the density of faint stars with $g-r \lesssim 1$, which will be sampled in future surveys such as the Rubin Observatory LSST and Euclid.
\item
Simulations over the entire DES-Y3 footprint based on our best-fit models were produced. Both simulations agree well with the data. Residual maps highlight many over-densities associated with globular clusters, dwarf galaxies, clouds, and streams in the DES footprint.
\item
We found a mass ratio between the thin and thick disks equal to $\simeq$ 13000:1, while the halo amounts to 1.1\% of the total MW stellar mass.
\end{itemize}
Future work with \mwf\ will include data from other wide-field surveys to extend the analysis to both the north and south Galactic hemispheres and will include improvements to the modeling for the Galactic halo (e.g., tri-axial models).

    \section*{Acknowledgements}

The authors are grateful to James Binney for many useful suggestions and comments. We thank Cecilia Mateu for the discussion about the location of the streams on the eastern part of the DES footprint.

Funding for the DES Projects has been provided by the U.S. Department of Energy, the U.S. National Science Foundation, the Ministry of Science and Education of Spain, the Science and Technology Facilities Council of the United Kingdom, the Higher Education Funding Council for England, the National Center for Supercomputing Applications at the University of Illinois at Urbana-Champaign, the Kavli Institute of Cosmological Physics at the University of Chicago, the Center for Cosmology and Astro-Particle Physics at the Ohio State University, the Mitchell Institute for Fundamental Physics and Astronomy at Texas A\&M University, Financiadora de Estudos e Projetos, Funda{\c c}{\~a}o Carlos Chagas Filho de Amparo {\`a} Pesquisa do Estado do Rio de Janeiro, Conselho Nacional de Desenvolvimento Cient{\'i}fico e Tecnol{\'o}gico and the Minist{\'e}rio da Ci{\^e}ncia, Tecnologia e Inova{\c c}{\~a}o, the Deutsche Forschungsgemeinschaft and the Collaborating Institutions in the Dark Energy Survey. 

The Collaborating Institutions are Argonne National Laboratory, the University of California at Santa Cruz, the University of Cambridge, Centro de Investigaciones Energ{\'e}ticas, Medioambientales y Tecnol{\'o}gicas-Madrid, the University of Chicago, University College London, the DES-Brazil Consortium, the University of Edinburgh, the Eidgen{\"o}ssische Technische Hochschule (ETH) Z{\"u}rich, Fermi National Accelerator Laboratory, the University of Illinois at Urbana-Champaign, the Institut de Ci{\`e}ncies de l'Espai (IEEC/CSIC), the Institut de F{\'i}sica d'Altes Energies, Lawrence Berkeley National Laboratory, the Ludwig-Maximilians Universit{\"a}t M{\"u}nchen and the associated Excellence Cluster Universe, the University of Michigan, the National Optical Astronomy Observatory, the University of Nottingham, The Ohio State University, the University of Pennsylvania, the University of Portsmouth, SLAC National Accelerator Laboratory, Stanford University, the University of Sussex, Texas A\&M University, and the OzDES Membership Consortium.

Based in part on observations at Cerro Tololo Inter-American Observatory, National Optical Astronomy Observatory, which is operated by the Association of 
Universities for Research in Astronomy (AURA) under a cooperative agreement with the National Science Foundation.

The DES data management system is supported by the National Science Foundation under Grant Numbers AST-1138766 and AST-1536171. The DES participants from Spanish institutions are partially supported by MINECO under grants AYA2015-71825, ESP2015-66861, FPA2015-68048, SEV-2016-0588, SEV-2016-0597, and MDM-2015-0509, some of which include ERDF funds from the European Union. IFAE is partially funded by the CERCA program of the Generalitat de Catalunya. Research leading to these results has received funding from the European Research Council under the European Union's Seventh Framework Program (FP7/2007-2013) including ERC grant agreements 240672, 291329, and 306478. We acknowledge support from the Australian Research Council Centre of Excellence for All-sky Astrophysics (CAASTRO), through project number CE110001020, and the Brazilian Instituto Nacional de Ci\^encia e Tecnologia (INCT) do e-Universe (CNPq grant 465376/2014-2).

This manuscript has been authored by Fermi Research Alliance, LLC under Contract No. DE-AC02-07CH11359 with the U.S. Department of Energy, Office of Science, Office of High Energy Physics. The United States Government retains and the publisher, by accepting the article for publication, acknowledges that the United States Government retains a non-exclusive, paid-up, irrevocable, worldwide license to publish or reproduce the published form of this manuscript, or allow others to do so, for United States Government purposes.

\bibliographystyle{mn2e}

\begin{thebibliography}{99}

\bibitem[DES Collaboration(2018)]{2018ApJS..239...18A} Abbott, T.~M.~C., Abdalla, F.~B., Allam, S., et al.\ 2018, \apjs, 239, 18 

\bibitem[Aihara et al.(2018)]{2018PASJ...70S...8A} Aihara, H., Armstrong, R., Bickerton, S., et al.\ 2018, \pasj, 70, S8 

\bibitem[Anders et al.(2014)]{2014A&A...564A.115A} Anders, F., Chiappini, C., Santiago, B.~X., et al.\ 2014, \aap, 564, A115 

\bibitem[Arnouts et al.(2001)]{2001A&A...379..740A} Arnouts, S., Vandame, B., Benoist, C., et al.\ 2001, \aap, 379, 740 

\bibitem[Balbinot et al.(2015)]{2015MNRAS.449.1129B} Balbinot, E., Santiago, B.~X., Girardi, L., et al.\ 2015, \mnras, 449, 1129 

\bibitem[Balbinot, et al.(2016)]{2016ApJ...820...58B} Balbinot E., et al., 2016, ApJ, 820, 58

\bibitem[Bahcall \& Soneira(1981)]{1981ApJS...47..357B} Bahcall, J.~N., \& Soneira, R.~M.\ 1981, \apjs, 47, 357 

\bibitem[Barmina et al.(2002)]{2002A&A...385..847B} Barmina, R., Girardi, L., \& Chiosi, C.\ 2002, \aap, 385, 847 

\bibitem[Bechtol et al.(2015)]{2015ApJ...807...50B} Bechtol, K., Drlica-Wagner, A., Balbinot, E., et al.\ 2015, \apj, 807, 50 

\bibitem[Bell, et al.(2008)]{2008ApJ...680..295B} Bell E.~F., et al., 2008, ApJ, 680, 295

\bibitem[Belokurov, et al.(2018)]{2018MNRAS.478..611B} Belokurov V., Erkal D., Evans N.~W., Koposov S.~E., Deason A.~J., 2018, MNRAS, 478, 611

\bibitem[Bennett \& Bovy(2019)]{2019MNRAS.482.1417B} Bennett, M., \& Bovy, J.\ 2019, \mnras, 482, 1417 

\bibitem[Bensby et al.(2003)]{2003A&A...410..527B} Bensby, T., Feltzing, S., \& Lundstr{\"o}m, I.\ 2003, \aap, 410, 527 

\bibitem[Bensby \& Feltzing(2010)]{2010IAUS..265..300B} Bensby, T., \& Feltzing, S.\ 2010, Chemical Abundances in the Universe: Connecting First Stars to Planets, 265, 300 

\bibitem[Bernard et al.(2016)]{2016MNRAS.463.1759B} Bernard, E.~J., Ferguson, A.~M.~N., Schlafly, E.~F., et al.\ 2016, \mnras, 463, 1759 

\bibitem[Bertelli et al.(1994)]{1994A&AS..106..275B} Bertelli, G., Bressan, A., Chiosi, C., Fagotto, F., \& Nasi, E.\ 1994, \aaps, 106, 275 

\bibitem[Binney et al.(1997)]{1997MNRAS.288..365B} Binney, J., Gerhard, O., \& Spergel, D.\ 1997, \mnras, 288, 365 

\bibitem[Binney \& Tremaine(2008)]{2008gady.book.....B} Binney, J., \& Tremaine, S.\ 2008, Galactic Dynamics: Second Edition, by James Binney and Scott Tremaine.~ISBN 978-0-691-13026-2 (HB).~Published by Princeton University Press, Princeton, NJ USA, 2008.,  

\bibitem[Bland-Hawthorn \& Gerhard(2016)]{2016ARA&A..54..529B} Bland-Hawthorn, J., \& Gerhard, O.\ 2016, \araa, 54, 529 

\bibitem[Blanton et al.(2017)]{2017AJ....154...28B} Blanton, M.~R., Bershady, M.~A., Abolfathi, B., et al.\ 2017, \aj, 154, 28 

\bibitem[Boeche et al.(2013)]{2013A&A...559A..59B} Boeche, C., Siebert, A., Piffl, T., et al.\ 2013, \aap, 559, A59 

\bibitem[Bournaud et al.(2009)]{2009ApJ...707L...1B} Bournaud, F., Elmegreen, B.~G., \& Martig, M.\ 2009, \apjl, 707, L1 

\bibitem[Bovy et al.(2016)]{2016ApJ...823...30B} Bovy, J., Rix, H.-W., Schlafly, E.~F., et al.\ 2016, \apj, 823, 30 

\bibitem[Brook et al.(2004)]{2004ApJ...612..894B} Brook, C.~B., Kawata, D., Gibson, B.~K., \& Freeman, K.~C.\ 2004, \apj, 612, 894 

\bibitem[Bullock \& Johnston(2005)]{2005ApJ...635..931B} Bullock J.~S., Johnston K.~V., 2005, ApJ, 635, 931

\bibitem[Burke et al.(2018)]{2018AJ....155...41B} Burke, D.~L., Rykoff, E.~S., Allam, S., et al.\ 2018, \aj, 155, 41 

\bibitem[Cabrera-Lavers et al.(2005)]{2005A&A...433..173C} Cabrera-Lavers, A., Garz{\'o}n, F., \& Hammersley, P.~L.\ 2005, \aap, 433, 173 

\bibitem[Casagrande et al.(2011)]{2011A&A...530A.138C} Casagrande, L., Sch{\"o}nrich, R., Asplund, M., et al.\ 2011, \aap, 530, A138 

\bibitem[Chabrier et al.(2000)]{2000ApJ...542..464C} Chabrier, G., Baraffe, I., Allard, F., \& Hauschildt, P.\ 2000, \apj, 542, 464 

\bibitem[Chabrier(2003)]{2003PASP..115..763C} Chabrier, G.\ 2003, \pasp, 115, 763 

\bibitem[Chiba \& Beers(2000)]{2000AJ....119.2843C} Chiba, M., \& Beers, T.~C.\ 2000, \aj, 119, 2843 

\bibitem[Courteau et al.(2011)]{2011ApJ...739...20C} Courteau, S., Widrow, L.~M., McDonald, M., et al.\ 2011, \apj, 739, 20 

\bibitem[Czekaj et al.(2014)]{2014A&A...564A.102C} Czekaj, M.~A., Robin, A.~C., Figueras, F., Luri, X., \& Haywood, M.\ 2014, \aap, 564, A102 

\bibitem[de Jong et al.(2010)]{2010ApJ...714..663D} de Jong, J.~T.~A., Yanny, B., Rix, H.-W., et al.\ 2010, \apj, 714, 663 

\bibitem[De Vicente, S{\'a}nchez \& Sevilla-Noarbe(2016)]{2016MNRAS.459.3078D} De Vicente J., S{\'a}nchez E., Sevilla-Noarbe I., 2016, MNRAS, 459, 3078

\bibitem[Deason et al.(2011)]{2011MNRAS.416.2903D} Deason, A.~J., Belokurov, V., \& Evans, N.~W.\ 2011, \mnras, 416, 2903 

\bibitem[Deason, et al.(2013)]{2013ApJ...763..113D} Deason A.~J., Belokurov V., Evans N.~W., Johnston K.~V., 2013, ApJ, 763, 113


\bibitem[Deason et al.(2018)]{2018ApJ...852..118D} Deason, A.~J., Belokurov, V., \& Koposov, S.~E.\ 2018, \apj, 852, 118 

\bibitem[Deason, et al.(2018)]{2018ApJ...862L...1D} Deason A.~J., Belokurov V., Koposov S.~E., Lancaster L., 2018, ApJL, 862, L1

\bibitem[Deason, Belokurov \& Sanders(2019)]{2019MNRAS.490.3426D} Deason A.~J., Belokurov V., Sanders J.~L., 2019, MNRAS, 490, 3426

\bibitem[DES Collaboration(2005)]{DES2005} DES Collaboration 2005, ArXiv e-prints, arXiv:astro-ph/0510346

\bibitem[Desai et al.(2012)]{2012ApJ...757...83D} Desai, S., Armstrong, R., Mohr, J.~J., et al.\ 2012, \apj, 757, 83 

\bibitem[Dolphin(2002)]{2002MNRAS.332...91D} Dolphin, A.~E.\ 2002, \mnras, 332, 91 

\bibitem[Drlica-Wagner et al.(2015)]{2015ApJ...813..109D} Drlica-Wagner, A., Bechtol, K., Rykoff, E.~S., et al.\ 2015, \apj, 813, 109 

\bibitem[Drlica-Wagner et al.(2018)]{2018ApJS..235...33D} Drlica-Wagner, A., Sevilla-Noarbe, I., Rykoff, E.~S., et al.\ 2018, \apjs, 235, 33 

\bibitem[Eggen et al.(1962)]{1962ApJ...136..748E} Eggen, O.~J., Lynden-Bell, D., \& Sandage, A.~R.\ 1962, \apj, 136, 748 

\bibitem[Eidelman et al.(2004)]{2004PhLB..592....1P} Eidelman, S., Hayes, K.~G., et al.\ 2004, Physics Letters B, 592, 1 

\bibitem[Einasto(1965)]{1965TrAlm...5...87E}Einasto, J.: 1965, {\it Trudy Astrofizicheskogo Instituta Alma-Ata} {\bf 5}, 87.

\bibitem[Fausti Neto et al.(2018)]{2018A&C....24...52F} Fausti Neto, A., da Costa, L.~N., Carnero, A., et al.\ 2018, Astronomy and Computing, 24, 52 

\bibitem[Feigelson and Babu(2012)]{Feigelson2012} Feigelson, E. D., Babu, G. J.\ 2012, Modern Statistical Methods for Astronomy: With R Applications, by Eric D. Feigelson and G. Jogesh Babu.~ISBN 978-1-139-53609-7 (HB).~Published by Cambridge University Press, UK, 2012.

\bibitem[Flaugher et al.(2015)]{2015AJ....150..150F} Flaugher, B., Diehl, H.~T., Honscheid, K., et al.\ 2015, \aj, 150, 150 

\bibitem[Fletcher(1987)]{1987Prmo.book.....F} Fletcher, R.\ 1987, Practical methods of optimization, by Robert Fletcher.~ISBN 978-0-471-91547-8 (HB).~Published by Wiley, Hoboken, NJ USA, 1987.

\bibitem[Foreman-Mackey et al.(2013)]{2013PASP..125..306F} Foreman-Mackey, D., Hogg, D.~W., Lang, D., \& Goodman, J.\ 2013, \pasp, 125, 306 

\bibitem[Fuhrmann(1998)]{1998A&A...338..161F} Fuhrmann, K.\ 1998, \aap, 338, 161 

\bibitem[Fuhrmann(2008)]{2008MNRAS.384..173F} Fuhrmann, K.\ 2008, \mnras, 384, 173 

\bibitem[Gaia Collaboration, et al.(2018)]{2018A&A...616A..10G} Gaia Collaboration, et al., 2018, A\&A, 616, A10

\bibitem[Gilmore \& Reid(1983)]{1983MNRAS.202.1025G} Gilmore, G., \& Reid, N.\ 1983, \mnras, 202, 1025 

\bibitem[Girardi et al.(2000)]{2000A&AS..141..371G} Girardi, L., Bressan, A., Bertelli, G., \& Chiosi, C.\ 2000, \aaps, 141, 371 

\bibitem[Girardi et al.(2002)]{2002A&A...391..195G} Girardi, L., Bertelli, G., Bressan, A., et al.\ 2002, \aap, 391, 195 

\bibitem[Girardi et al.(2005)]{2005A&A...436..895G} Girardi, L., Groenewegen, M.~A.~T., Hatziminaoglou, E., \& da Costa, L.\ 2005, \aap, 436, 895 

\bibitem[Girardi et al.(2010)]{2010ApJ...724.1030G} Girardi, L., Williams, B.~F., Gilbert, K.~M., et al.\ 2010, \apj, 724, 1030 

\bibitem[Girardi et al.(2012)]{2012ASSP...26..165G} Girardi, L., Barbieri, M., Groenewegen, M.~A.~T., et al.\ 2012, \assp, 26, 165 

\bibitem[Gravity Collaboration et al.(2018)]{2018A&A...615L..15G} Gravity Collaboration, Abuter, R., Amorim, A., et al.\ 2018, \aap, 615, L15 

\bibitem[Grillmair \& Dionatos(2006)]{2006ApJ...643L..17G} Grillmair, C.~J., \& Dionatos, O.\ 2006, \apjl, 643, L17 

\bibitem[Groenewegen et al.(2002)]{2002A&A...392..741G} Groenewegen, M.~A.~T., Girardi, L., Hatziminaoglou, E., et al.\ 2002, \aap, 392, 741 

\bibitem[Gschwend et al.(2018)]{2018A&C....25...58G} Gschwend, J., Rossel, A.~C., Ogando, R.~L.~C., et al.\ 2018, Astronomy and Computing, 25, 58 

\bibitem[Harris(1996)]{1996AJ....112.1487H} Harris, W.~E.\ 1996, \aj, 112, 1487 

\bibitem[Helmi(2016)]{2016IAUS..317..228H} Helmi, A.\ 2016, The General Assembly of Galaxy Halos: Structure, Origin and Evolution, 317, 228 

\bibitem[Helmi et al.(2018)]{2018Natur.563...85H} Helmi, A., Babusiaux, C., Koppelman, H.~H., et al.\ 2018, \nat, 563, 85 

\bibitem[Hernitschek, et al.(2018)]{2018ApJ...859...31H} Hernitschek N., et al., 2018, \apj, 859, 31

\bibitem[Holmberg \& Flynn(2004)]{2004MNRAS.352..440H} Holmberg J., Flynn C., 2004, MNRAS, 352, 440

\bibitem[Hopkins et al.(2014)]{2014MNRAS.445..581H} Hopkins, P.~F., Kere{\v s}, D., O{\~n}orbe, J., et al.\ 2014, \mnras, 445, 581 

\bibitem[Ibata et al.(2019)]{2019ApJ...872..152I} Ibata, R.~A., Malhan, K., \& Martin, N.~F.\ 2019, \apj, 872, 152 

\bibitem[Juri{\'c} et al.(2008)]{2008ApJ...673..864J} Juri{\'c}, M., Ivezi{\'c}, {\v Z}., Brooks, A., et al.\ 2008, \apj, 673, 864 

\bibitem[Lindegren et al.(2016)]{2016A&A...595A...4L} Lindegren, L., Lammers, U., Bastian, U., et al.\ 2016, \aap, 595, A4 

\bibitem[Kim \& Jerjen(2015)]{2015ApJ...808L..39K} Kim, D., \& Jerjen, H.\ 2015, \apjl, 808, L39 

\bibitem[Kleinman et al.(2004)]{2004ApJ...607..426K} Kleinman, S.~J., Harris, H.~C., Eisenstein, D.~J., et al.\ 2004, \apj, 607, 426 

\bibitem[Koposov et al.(2014)]{2014MNRAS.442L..85K} Koposov, S.~E., Irwin, M., Belokurov, V., et al.\ 2014, \mnras, 442, L85 

\bibitem[Koposov et al.(2015)]{2015ApJ...805..130K} Koposov, S.~E., Belokurov, V., Torrealba, G., \& Evans, N.~W.\ 2015, \apj, 805, 130 

\bibitem[Kroupa \& Weidner(2003)]{2003ApJ...598.1076K} Kroupa, P., \& Weidner, C.\ 2003, \apj, 598, 1076 

\bibitem[Kroupa(2001)]{2001MNRAS.322..231K} Kroupa, P.\ 2001, \mnras, 322, 231 

\bibitem[Li et al.(2016)]{2016ApJ...817..135L} Li, T.~S., Balbinot, E., Mondrik, N., et al.\ 2016, \apj, 817, 135 

\bibitem[Licquia \& Newman(2013)]{2013AAS...22125411L} Licquia, T., \& Newman, J.\ 2013, American Astronomical Society Meeting Abstracts \#221, 221, 254.11 

\bibitem[Licquia \& Newman(2015)]{2015ApJ...806...96L} Licquia, T.~C., \& Newman, J.~A.\ 2015, \apj, 806, 96 

\bibitem[Loebman et al.(2011)]{2011ApJ...737....8L} Loebman, S.~R., Ro{\v s}kar, R., Debattista, V.~P., et al.\ 2011, \apj, 737, 8 

\bibitem[LSST Science Collaboration et al.(2009)]{2009arXiv0912.0201L} LSST Science Collaboration, Abell, P.~A., Allison, J., et al.\ 2009, arXiv:0912.0201 

\bibitem[Luque et al.(2018)]{2018MNRAS.478.2006L} Luque, E., Santiago, B., Pieres, A., et al.\ 2018, \mnras, 478, 2006 

\bibitem[Lynga(1982)]{1982A&A...109..213L} Lyng\aa, G.\ 1982, \aap, 109, 213 

\bibitem[Mackey et al.(2018)]{2018ApJ...858L..21M} Mackey, D., Koposov, S., Da Costa, G., et al.\ 2018, \apjl, 858, L21 

\bibitem[Ma{\'{\i}}z-Apell{\'a}niz(2001)]{2001AJ....121.2737M} Ma{\'{\i}}z-Apell{\'a}niz, J.\ 2001, \aj, 121, 2737 

\bibitem[Majewski et al.(2003)]{2003ApJ...599.1082M} Majewski, S.~R., Skrutskie, M.~F., Weinberg, M.~D., \& Ostheimer, J.~C.\ 2003, \apj, 599, 1082 

\bibitem[Majewski et al.(2016)]{2016AN....337..863M} Majewski, S.~R., APOGEE Team, \& APOGEE-2 Team 2016, Astronomische Nachrichten, 337, 863

\bibitem[Mackereth \& Bovy(2020)]{2020MNRAS.492.3631M} Mackereth J.~T., Bovy J., 2020, MNRAS, 492, 3631


\bibitem[Marigo \& Girardi(2007)]{2007A&A...469..239M} Marigo, P., \& Girardi, L.\ 2007, \aap, 469, 239 

\bibitem[Marigo et al.(2017)]{2017ApJ...835...77M} Marigo P., et al., 2017, \apj, 835, 77 

\bibitem[Mateu et al.(2018)]{2018MNRAS.474.4112M} Mateu, C., Read, J.~I., \& Kawata, D.\ 2018, \mnras, 474, 4112 

\bibitem[Mateu(2017)]{2017ascl.soft11010M} Mateu, C.\ 2017, Astrophysics Source Code Library, ascl:1711.010 

\bibitem[McConnachie(2012)]{2012AJ....144....4M} McConnachie, A.~W.\ 2012, \aj, 144, 4 

\bibitem[McMillan(2011)]{2011MNRAS.414.2446M} McMillan, P.~J.\ 2011, \mnras, 414, 2446 

\bibitem[Mehta, et al.{2019}]{2019yCat..22350036M} Mehta V., et al., 2019, yCat, J/ApJS/235/36

\bibitem[Merritt et al.(2006)]{2006AJ....132.2685M} Merritt D., Graham A.~W., Moore B., Diemand J., Terzi{\'c} B., 2006, \aj, 132, 2685

\bibitem[Minchev et al.(2015)]{2015ApJ...804L...9M} Minchev, I., Martig, M., Streich, D., et al.\ 2015, \apjl, 804, L9 

\bibitem[Morganson et al.(2018)]{2018PASP..130g4501M} Morganson, E., Gruendl, R.~A., Menanteau, F., et al.\ 2018, \pasp, 130, 074501 

\bibitem[Mohr et al.(2008)]{2008SPIE.7016E..0LM} Mohr, J.~J., Adams, D., Barkhouse, W., et al.\ 2008, \procspie, 7016, 70160L 

\bibitem[Mohr et al.(2012)]{2012SPIE.8451E..0DM} Mohr, J.~J., Armstrong, R., Bertin, E., et al.\ 2012, \procspie, 8451, 84510D 

\bibitem[Nataf et al.(2013)]{2013ApJ...769...88N} Nataf, D.~M., Gould, A., Fouqu{\'e}, P., et al.\ 2013, \apj, 769, 88 

\bibitem[Newberg et al.(2002)]{2002ApJ...569..245N} Newberg, H.~J., Yanny, B., Rockosi, C., et al.\ 2002, \apj, 569, 245 

\bibitem[Ngeow et al.(2006)]{2006SPIE.6270E..23N} Ngeow, C., Mohr, J.~J., Alam, T., et al.\ 2006, \procspie, 6270, 627023 

\bibitem[Nidever et al.(2019)]{2019ApJ...874..118N} Nidever, D.~L., Olsen, K., Choi, Y., et al.\ 2019, \apj, 874, 118 

\bibitem[Odenkirchen et al.(2003)]{2003AJ....126.2385O} Odenkirchen, M., Grebel, E.~K., Dehnen, W., et al.\ 2003, \aj, 126, 2385 

\bibitem[Osmer et al.(1998)]{1998ApJS..119..189O} Osmer, P.~S., Kennefick, J.~D., Hall, P.~B., \& Green, R.~F.\ 1998, \apjs, 119, 189 

\bibitem[Pasetto et al.(2018)]{2018ApJ...860..120P} Pasetto, S., Grebel, E.~K., Chiosi, C., et al.\ 2018, \apj, 860, 120 

\bibitem[Paxton et al.(2011)]{2011ApJS..192....3P} Paxton, B., Bildsten, L., Dotter, A., et al.\ 2011, \apjs, 192, 3 

\bibitem[Perryman et al.(1997)]{1997A&A...323L..49P} Perryman, M.~A.~C., Lindegren, L., Kovalevsky, J., et al.\ 1997, \aap, 323, L49 

\bibitem[Pila-D{\'\i}ez, et al.(2015)]{2015A&A...579A..38P} Pila-D{\'\i}ez B., de Jong J.~T.~A., Kuijken K., van der Burg R.~F.~J., Hoekstra H., 2015, A\&A, 579, A38

\bibitem[Pieres et al.(2017)]{2017MNRAS.468.1349P} Pieres, A., Santiago, B.~X., Drlica-Wagner, A., et al.\ 2017, \mnras, 468, 1349 

\bibitem[Portail et al.(2017)]{2017MNRAS.465.1621P} Portail, M., Gerhard, O., Wegg, C., \& Ness, M.\ 2017, \mnras, 465, 1621 

\bibitem[Rana \& Basu(1992)]{1992A&A...265..499R} Rana, N.~C., \& Basu, S.\ 1992, \aap, 265, 499 

\bibitem[Reddy et al.(2006)]{2006MNRAS.367.1329R} Reddy, B.~E., Lambert, D.~L., \& Allende Prieto, C.\ 2006, \mnras, 367, 1329 

\bibitem[Rocha-Pinto et al.(2000)]{2000A&A...358..850R} Rocha-Pinto, H.~J., Maciel, W.~J., Scalo, J., \& Flynn, C.\ 2000, \aap, 358, 850 

\bibitem[Ryan \& Norris(1991)]{1991AJ....101.1865R} Ryan, S.~G., \& Norris, J.~E.\ 1991, \aj, 101, 1865 

\bibitem[Sarajedini et al.(2007)]{2007AJ....133.1658S} Sarajedini, A., Bedin, L.~R., Chaboyer, B., et al.\ 2007, \aj, 133, 1658 

\bibitem[Sartoris et al.(2016)]{2016MNRAS.459.1764S} Sartoris, B., Biviano, A., Fedeli, C., et al.\ 2016, \mnras, 459, 1764 

\bibitem[Schlegel et al.(1998)]{1998ApJ...500..525S} Schlegel, D.~J., Finkbeiner, D.~P., \& Davis, M.\ 1998, \apj, 500, 525 

\bibitem[Sch{\"o}nrich \& Binney(2009)]{2009MNRAS.396..203S} Sch{\"o}nrich, R., \& Binney, J.\ 2009, \mnras, 396, 203 

\bibitem[Sesar et al.(2011)]{2011ApJ...731....4S} Sesar, B., Juri{\'c}, M., \& Ivezi{\'c}, {\v Z}.\ 2011, \apj, 731, 4 

\bibitem[Sevilla et al.(2011)]{2011arXiv1109.6741S} Sevilla, I., Armstrong, R., Bertin, E., et al.\ 2011, arXiv:1109.6741 

\bibitem[Sevilla-Noarbe et al.(2018)]{2018MNRAS.481.5451S} Sevilla-Noarbe, I., Hoyle, B., March{\~a}, M.~J., et al.\ 2018, \mnras, 481, 5451 

\bibitem[Sheldon(2015)]{2015ascl.soft08008S} Sheldon, E.\ 2015, Astrophysics Source Code Library, ascl:1508.008 

\bibitem[Skrutskie et al.(2006)]{2006AJ....131.1163S} Skrutskie, M.~F., Cutri, R.~M., Stiening, R., et al.\ 2006, \aj, 131, 1163 

\bibitem[Sharma et al.(2011)]{2011ApJ...730....3S} Sharma, S., Bland-Hawthorn, J., Johnston, K.~V., \& Binney, J.\ 2011, \apj, 730, 3 

\bibitem[Shipp et al.(2018)]{2018AAS...23121205S} Shipp, N., Drlica-Wagner, A., Balbinot, E., \& DES Collaboration 2018, \apj, 862, 114 

\bibitem[Slater et al.(2016)]{2016ApJ...832..206S} Slater, C.~T., Nidever, D.~L., Munn, J.~A., Bell, E.~F., \& Majewski, S.~R.\ 2016, \apj, 832, 206 

\bibitem[Spada et al.(2013)]{2013ApJ...776...87S} Spada, F., Demarque, P., Kim, Y.-C., \& Sills, A.\ 2013, \apj, 776, 87 

\bibitem[Steinmetz(2012)]{2012AN....333..523S} Steinmetz, M.\ 2012, Astronomische Nachrichten, 333, 523 

\bibitem[Stoughton et al.(2002)]{2002AJ....123..485S} Stoughton, C., Lupton, R.~H., Bernardi, M., et al.\ 2002, \aj, 123, 485 

\bibitem[Swanson et al.(2008)]{2008MNRAS.387.1391S} Swanson, M.~E.~C., Tegmark, M., Hamilton, A.~J.~S., \& Hill, J.~C.\ 2008, \mnras, 387, 1391 

\bibitem[Torrealba et al.(2015)]{2015MNRAS.446.2251T} Torrealba, G., Catelan, M., Drake, A.~J., et al.\ 2015, \mnras, 446, 2251 

\bibitem[VandenBerg et al.(2006)]{2006ApJS..162..375V} VandenBerg, D.~A., Bergbusch, P.~A., \& Dowler, P.~D.\ 2006, \apjs, 162, 375 

\bibitem[Vanhollebeke et al.(2009)]{2009A&A...498...95V} Vanhollebeke, E., Groenewegen, M.~A.~T., \& Girardi, L.\ 2009, \aap, 498, 95 

\bibitem[Villalobos \& Helmi(2008)]{2008MNRAS.391.1806V} Villalobos, {\'A}., \& Helmi, A.\ 2008, \mnras, 391, 1806 

\bibitem[Watkins et al.(2009)]{2009MNRAS.398.1757W} Watkins, L.~L., Evans, N.~W., Belokurov, V., et al.\ 2009, \mnras, 398, 1757 

\bibitem[Wood \& Mao(2005)]{2005MNRAS.362..945W} Wood, A., \& Mao, S.\ 2005, \mnras, 362, 945 

\bibitem[Xue et al. (2015)]{2015ApJ...809...144} Xue, X.-X., Rix, H.-W., Ma, Z., et al.\ 2015, \apj, 809, 144

\bibitem[Yanny et al.(2009)]{2009ApJ...700.1282Y} Yanny, B., Newberg, H.~J., Johnson, J.~A., et al.\ 2009, \apj, 700, 1282 

\bibitem[Zoccali et al.(2003)]{2003A&A...399..931Z} Zoccali, M., Renzini, A., Ortolani, S., et al.\ 2003, \aap, 399, 931 

\end{thebibliography}
{}

\section*{Affiliations}
{\small\it
\noindent
$^{1}$ Laborat\'orio Interinstitucional de e-Astronomia - LIneA, Rua Gal. Jos\'e Cristino 77, Rio de Janeiro, RJ - 20921-400, Brazil \\
$^{2}$ Observat\'orio Nacional, Rua Gal. Jos\'e Cristino 77, Rio de Janeiro, RJ - 20921-400, Brazil \\
$^{3}$ Osservatorio Astronomico di Padova, INAF, Vicolo dell'Osservatorio 5, I-35122 Padova, Italy \\
$^{4}$ Kapteyn Astronomical Institute, University of Groningen, Landleven 12, 9747 AD Groningen, The Netherlands \\
$^{5}$ Instituto de F\'\i sica, UFRGS, Caixa Postal 15051, Porto Alegre, RS - 91501-970, Brazil \\
$^{6}$ Centro de Investigaciones Energ\'eticas, Medioambientales y Tecnol\'ogicas (CIEMAT), Madrid, Spain \\
$^{7}$ George P. and Cynthia Woods Mitchell Institute for Fundamental Physics and Astronomy, and Department of Physics and Astronomy, Texas A\&M University, College Station, TX 77843, USA \\
$^{8}$ LSST 933 North Cherry Avenue, Tucson, Arizona 85721, USA \\
$^{9}$ Koninklijke Sterrenwacht van Belgi\"e, Ringlaan 3, B–1180 Brussels, Belgium \\
$^{10}$ Fermi National Accelerator Laboratory, P. O. Box 500, Batavia, IL 60510, USA \\
$^{11}$ Kavli Institute for Cosmological Physics, University of Chicago, Chicago, IL 60637, USA \\
$^{12}$ Department of Physics, ETH Zurich, Wolfgang-Pauli-Strasse 16, CH-8093 Zurich, Switzerland \\
$^{13}$ Instituto de Fisica Teorica UAM/CSIC, Universidad Autonoma de Madrid, 28049 Madrid, Spain \\
$^{14}$ CNRS, UMR 7095, Institut d'Astrophysique de Paris, F-75014, Paris, France \\
$^{15}$ Sorbonne Universit\'es, UPMC Univ Paris 06, UMR 7095, Institut d'Astrophysique de Paris, F-75014, Paris, France \\
$^{16}$ Department of Physics \& Astronomy, University College London, Gower Street, London, WC1E 6BT, UK \\
$^{17}$ Kavli Institute for Particle Astrophysics \& Cosmology, P. O. Box 2450, Stanford University, Stanford, CA 94305, USA \\
$^{18}$ SLAC National Accelerator Laboratory, Menlo Park, CA 94025, USA \\
$^{19}$ Department of Astronomy, University of Illinois at Urbana-Champaign, 1002 W. Green Street, Urbana, IL 61801, USA \\
$^{20}$ National Center for Supercomputing Applications, 1205 West Clark St., Urbana, IL 61801, USA \\
$^{21}$ Institut de F\'{\i}sica d'Altes Energies (IFAE), The Barcelona Institute of Science and Technology, Campus UAB, 08193 Bellaterra (Barcelona) Spain \\
$^{22}$ Department of Physics, IIT Hyderabad, Kandi, Telangana 502285, India \\
$^{23}$ Department of Astronomy/Steward Observatory, University of Arizona, 933 North Cherry Avenue, Tucson, AZ 85721-0065, USA \\
$^{24}$ Jet Propulsion Laboratory, California Institute of Technology, 4800 Oak Grove Dr., Pasadena, CA 91109, USA \\
$^{25}$ Institut d'Estudis Espacials de Catalunya (IEEC), 08034 Barcelona, Spain \\
$^{26}$ Institute of Space Sciences (ICE, CSIC),  Campus UAB, Carrer de Can Magrans, s/n,  08193 Barcelona, Spain \\
$^{27}$ Department of Astronomy, University of Michigan, Ann Arbor, MI 48109, USA \\
$^{28}$ Department of Physics, University of Michigan, Ann Arbor, MI 48109, USA \\
$^{29}$ Department of Physics, Stanford University, 382 Via Pueblo Mall, Stanford, CA 94305, USA \\
$^{30}$ Santa Cruz Institute for Particle Physics, Santa Cruz, CA 95064, USA \\
$^{31}$ Center for Cosmology and Astro-Particle Physics, The Ohio State University, Columbus, OH 43210, USA \\
$^{32}$ Department of Physics, The Ohio State University, Columbus, OH 43210, USA \\
$^{33}$ Harvard-Smithsonian Center for Astrophysics, Cambridge, MA 02138, USA \\
$^{34}$ Australian Astronomical Optics, Macquarie University, North Ryde, NSW 2113, Australia \\
$^{35}$ Instituci\'o Catalana de Recerca i Estudis Avan\c{c}ats, E-08010 Barcelona, Spain \\
$^{36}$ Department of Astrophysical Sciences, Princeton University, Peyton Hall, Princeton, NJ 08544, USA \\
$^{37}$ Brookhaven National Laboratory, Bldg 510, Upton, NY 11973, USA \\
$^{38}$ School of Physics and Astronomy, University of Southampton,  Southampton, SO17 1BJ, UK \\
$^{39}$ Brandeis University, Physics Department, 415 South Street, Waltham MA 02453 \\
$^{40}$ Instituto de F\'isica Gleb Wataghin, Universidade Estadual de Campinas, 13083-859, Campinas, SP, Brazil \\
$^{41}$ Computer Science and Mathematics Division, Oak Ridge National Laboratory, Oak Ridge, TN 37831 \\
$^{42}$ Institute of Cosmology and Gravitation, University of Portsmouth, Portsmouth, PO1 3FX, UK \\
$^{43}$ Argonne National Laboratory, 9700 South Cass Avenue, Lemont, IL 60439, USA \\
$^{44}$ Cerro Tololo Inter-American Observatory, National Optical Astronomy Observatory, Casilla 603, La Serena, Chile \\
}
\label{lastpage}
\appendix




\section{MWFitting pipeline inputs}
This appendix describes input parameters of the pipeline when submitting \textsc{MWFitting} through the science portal, intended to guide LIneA users. Table \ref{inputpars} lists the name, description, standard configuration and units of the parameters that the user should use to reproduce our results. These parameters are not related to the components of the MW models, but needed to run the code.

\begin{table*}
\begin{center}
\def\arraystretch{1.}%
\caption{The main parameters to run \mwf\,  pipeline.}
\setlength{\tabcolsep}{6pt}
\begin{tabular}{l l l r}
\hline
Parameter name &  Description & Std. configuration & Unit \\
\hline
\multicolumn{4}{c}{\textbf{Input data}} \\
NSIDE & Footprint map granularity & 1024 & HEALPix Nside \\
Input as simulation & Build and fit mock catalogues & False & - \\
Random factor & Range to multiply initial values & 2 & - \\
 & (for ‘input as simulation’ method) & & \\
Seed for input & Seed to generate random numbers & 0 & - \\
 & and multiply random factor & & \\
 \hline
\multicolumn{4}{c}{\bf{Mock catalogues}} \\    
 Build mock catalogue? & Build a mock catalogue & False & - \\
 & following the best-fitting parameters & & \\
 \hline
\multicolumn{4}{c}{\bf{Hess Diagrams}} \\
Nside  & Size of cells for HD  & 16  & HEALPix Nside \\
Minimum area  & minimum coverage of the cell & 0.9 & Cell area \\
Global seed  & Global seed for fields choice  & 1  & -  \\
Magnitude range  & min, max, step  & 17.0;21.0;0.1  & mag  \\
Colour range & min, max, step & 0.0;0.8;0.1  & mag  \\
Cell counts & Number of fields to be fitted & 10 & Field \\
\hline
\multicolumn{4}{c}{\bf{Filters}} \\
Streams  & Filter streams  & True  & - \\
Minimum Galactic latitude  & Lowest absolute value for $b$  & 30  & degree \\
\hline
\multicolumn{4}{c}{\bf{Optimise}} \\
Overfactor  & Oversampling the models  & 20  & area in cell \\
\hline
\end{tabular}
\label{inputpars}
\end{center}
\end{table*}

\end{document}